\documentclass[a4paper,11pt]{article}
\usepackage{amssymb}
\usepackage{amsfonts}
\usepackage{float}
\usepackage{graphicx}
\usepackage{color,subfigure}

\newcommand{\lsim}{\mathrel{\lower0.55ex\hbox{$\mathchar"3218$}\mkern-14mu\raise0.55ex\hbox{$<$}}}
\newcommand{\gsim}{\mathrel{\lower0.55ex\hbox{$\mathchar"3218$}\mkern-14mu\raise0.55ex\hbox{$>$}}}

\begin{document}

\title{A moving boundary model motivated by electric breakdown:
II. Initial value problem}

\author{C.-Y. Kao$^1$, F. Brau$^{2,3}$, U. Ebert$^{2,4}$, L. Sch\"afer$^5$,
S. Tanveer$^1$ \\
~\\
$^1$ Department of Mathematics,\\ The Ohio State University, OH 43210, USA,\\
$^2$ Centrum Wiskunde \& Informatica (CWI), P.O.Box 94079, \\ 1090GB Amsterdam, The Netherlands,\\
$^3$ Groupe de Physique Nucleaire Th\'eorique, \\Universit\'e de Mons-Hainaut, \\
Acad\'emie universitaire Wallonie-Bruxelles, \\ Place du Parc 20, 7000 Mons, Belgium\\
$^4$ Department of Physics, Eindhoven University of Technology, \\Eindhoven, The Netherlands,\\
$^5$ Fachbereich Physik, Universit\"at Duisburg-Essen,\\ Lotharstr.~1, 47048 Duisburg, Germany.}

\date{\today}

\maketitle

\begin{abstract} {
An interfacial approximation of the streamer stage in the evolution of
sparks and lightning can be formulated as a Laplacian growth model
regularized by a 'kinetic undercooling' boundary condition. Using this
model we study both the linearized and the full nonlinear evolution of
small perturbations of a uniformly translating circle. Within the linear
approximation analytical and numerical results show that perturbations
are advected to the back of the circle, where they decay. An initially
analytic interface stays analytic for all finite times, but
singularities from outside the physical region approach the interface
for $t\rightarrow\infty$, which results in some anomalous relaxation at
the back of the circle. For the nonlinear evolution numerical results
indicate that the circle is the asymptotic attractor for small
perturbations, but larger perturbations may lead to branching. We also
present results for more general initial shapes, which demonstrate that
regularization by kinetic undercooling cannot guarantee smooth
interfaces globally in time.
}
\\~\\
PACS: 47.54.-r
\\~\\
Keywords: moving boundary, kinetic undercooling regularization, initial value problem, Laplacian
instability, electric breakdown
\end{abstract}



\renewcommand {\thesubsection}{\arabic{section}\Alph {subsection}}

\section{Introduction} \label{introduction}

Propagating fronts in Laplacian growth occur naturally in quite a number of physical problems including viscous fingering~\cite{SaffmanTaylor, bens86, Homsy, Bensimon_etal, Tanveer00}, electro-chemical growth, dendritic crystal growth for small undercooling~\cite{BenAmar, Pelce, KesslerKoplikLevine}, and void migration in a conductor~\cite{mah-96, amar-99, Cummingsetal}. More recently, it has been shown that this class of problems includes the 'streamer' stage of electric breakdown~\cite{eber96, man02, eber06, brau08, meul04, meul05, eber07, PartI}, which will be described below. A central issue in these problems is the stability of curved fronts. In a limiting case, most of these models reduce to the classic Saffman-Taylor problem~\cite{SaffmanTaylor}, which is known to be ill-posed \cite{Howison, FokasTanveer}. Numerical as well as formal asymptotic results~\cite{Bensimon_etal, KesslerKoplikLevine, Tanveer00, TanveerSaffman} suggest that one branch of steadily propagating finger or bubble solutions in a Hele-Shaw cell is stabilized by surface tension regularization, though only recently some mathematically rigorous results~\cite{Ye-Tanveer,Ye-Tanveer2} are available to justify nonlinear stability to small disturbances in the special case of a nearly circular bubble. Besides surface tension, other regularizations~\cite{mah-96, amar-99, Cummingsetal, Reissig99} have also been analyzed. In the present paper we study both the linear and the nonlinear initial value problem for one such regularization, in particular, the stability of a steadily propagating circular shape. This regularization is called kinetic undercooling in the crystal growth context\footnote{In crystal growth, kinetic undercooling is likely to be the more important regularization compared to the Gibbs-Thompson effect for large undercooling where the Laplacian growth model becomes questionable. This happens when the time scale over which the interface evolves becomes comparable to the time scale of heat diffusion.}, but has a different physical interpretation for streamers.

During the streamer stage of electric breakdown the discharge paves its way through a nonconducting medium, leaving behind a weakly ionized conducting channel. The basic growth mechanism is impact ionization due to electrons strongly accelerated in the local electric field. In a sufficiently strong field, a thin space charge layer forms around the head of the streamer. This layer screens the field in the inner ionized region to a very low level, and the growth of the streamer is driven by the electrons moving and multiplying in the strong self-enhanced field ahead of the curved ionization front.

For sufficiently strong external fields, the thickness $\ell$ of the electron layer is small compared to the
radius $R$ of the streamer head. Therefore Lozansky and Firsov suggested (mainly in the Russian literature,
but also in \cite{LF}) that this layer can be modeled as an interface separating the ionized from the
non-ionized region. Probably, the idea is even older, since a similar concept was already proposed by S\"ammer in the German literature in 1933~\cite{Sammer}. However, a deeper study of the implications of this concept
started only later~\cite{eber96, man02, eber06, man04, gianne08, ST08} where the problem is placed in the
context of other Laplacian moving boundary problems. The validity of the moving boundary approximation for
negative streamers is discussed in~\cite{brau08} for simple gases like pure nitrogen or argon, and
in~\cite{LuRaEb08} for air. The dimensional analysis and the proposed regularization mechanism of the moving
boundary problem for negative streamer ionization fronts are discussed in detail in our previous
papers~\cite{meul04, meul05, eber07, PartI}.

In dimensionless form, the model is defined as follows. The normal velocity $v_{n}$ of the interface is
given by the drift velocity ${\bf v}$ of the electrons, which is proportional to the local electro-static
field ${\bf E}=-\nabla \varphi$. In appropriate units it takes the dimensionless form:
\begin{equation} \label{Achat}
v_{n} = - {\bf n} \cdot {\bf E}^+ \,,
\end{equation}
where the super-script $^+$ denotes the limiting value as the interface is approached from the exterior
(the non-ionized region) and ${\bf n}$ is the outward normal on the interface. Outside the streamer the electric potential $\varphi$ obeys the Laplace equation:
\begin{equation} \label{Amazonit}
\Delta \varphi = 0\,.
\end{equation}
An analytical and numerical analysis of the underlying physical model formulated in terms of partial
differential equations for charge densities and field suggests the interfacial condition
\begin{equation} \label{Amethyst}
\varphi^+ = \epsilon \, {\bf n} \cdot (\nabla \varphi)^+ ,
\end{equation}
where
\begin{equation} \label {Ametrin}
\epsilon = \frac{\ell}{R} \,.
\end{equation}
Far from the streamer, the electric field tends to a constant\footnote{ A correction of order $O(1/|{\bf
x}|)$ to the electric field {\bf E} can occur only if the streamer carries a net electric charge. We here
concentrate on the analysis of streamers that are globally electrically neutral.}
\begin{equation} \label{Antimonit}
{\bf E} = - \nabla \varphi \to - \ \hat{\bf x} + o(1/|{\bf x}|) \,,
\end{equation}
where $\hat{\bf x}$ is the unit vector in $x$-direction. Eqs.~(\ref{Achat})--(\ref{Antimonit}) define our
model.

In two dimensions, a simple solution to the free boundary problem posed by this model takes the form of a
uniformly translating circle. Our previous work in \cite{meul05, eber07, PartI} and the present paper are
primarily concerned with the linear and nonlinear stability of this solution to small perturbations. It is
to be noted that the circular shape differs from an actual streamer shape. However, the front half of a
circle roughly resembles the shape of the front part of a streamer. Since growth of disturbances is found to
be most pronounced in this advancing part of the interface, we expect stability features found here to be
qualitatively relevant for an actual streamer and more generally for curved fronts.

In the special case $\epsilon=1$, the linearized evolution of small perturbations can be determined exactly
in our model~\cite{meul05, eber07}. The case of general $\epsilon>0$ is treated in part I~\cite{PartI} of
this series of papers and in the present manuscript. In~\cite{PartI}, we discussed the spectrum of the
linear operator which results from the linear stability analysis of the circular solution. Restricting
ourselves to an appropriate space of analytic perturbations we found a pure point spectrum. Asymptotically
in time, except for the trivial translation mode, all eigenmodes were found to decay exponentially in time.
These eigenmodes are singular at the back of the bubble; nonetheless, as evidenced in the present paper,
this singularity is not reflected by the actual linear evolution near the bubble back. The usual asymptotic
form of the solution for large time: $\sum_{\lambda} e^{\lambda t} \beta_{\lambda} $, where $\beta_\lambda$
is the eigenfunction corresponding to the eigenvalue $\lambda$, fails in a neighborhood of the rear of the
bubble, though it holds elsewhere.

In the present paper, we consider the initial value problem. For the linearized evolution, analytical results are obtained in the limit of strongly localized disturbances of the circle. Also the large time behavior of general perturbations can be studied analytically. Numerical calculations confirm these results. Together with the eigenvalue analysis of the first
paper~\cite{PartI}, clear evidence of linear stability is presented. The full nonlinear evolution of a
perturbed interface is calculated numerically. Our results suggest that, similar to linear evolution, small
enough perturbations of a circular bubble grow in the front part of the bubble, but eventually decay as
interfacial distortions advect to the bubble rear. Nonetheless, when $\epsilon$ is small but nonzero, the
large transients in the linear regime make nonlinearity important even when the initial perturbation is
exponentially small in $\epsilon$. Furthermore, when the perturbations are larger, the circle is no longer
an attractor of the dynamics and the propagating structure branches. For general initial shapes, we give
some numerical evidence that the undercooling regularization condition can not guarantee a smooth interface
globally in time. For some initial conditions, the interface tends to develop a sharp corner in the back. Other initial conditions lead to the separation of the moving body into two parts.

This paper is organized as follows. In section~\ref{equations} we present equations derived earlier in a
conformal map setting. Section~\ref{analysis} is devoted to the linear evolution of perturbations of the
circle. Subsection~\ref{previous results} recalls previous results, and in subsection~\ref{stronglyrigor} we present rigorous results on the growth of a strongly localized
perturbation. We continue the discussion of localized perturbations in subsection \ref{strongly} and explain
at an intuitive level how strongly localized perturbations are generically advected to the rear of the
circle, increasing in amplitude in the front-half before decreasing in the back half.
Mathematically, the advection is described by a one-parameter family of conformal maps which is a subgroup
of the automorphisms of the unit disk. The important role of this subgroup has been previously established
for the exactly solvable case $\epsilon = 1$ \cite{meul05, eber07}. In subsection~\ref{asymptotic} we
discuss the anomalous behavior found at the back of the circle in the large time limit. In
subsection~\ref{smoothness} we give arguments indicating that an initially analytic interface stays analytic
for all finite times, but singularities initially outside the physical region of interest approach
the back of the circle for $t \to \infty$. Provided the perturbation for $t \to \infty$ stays analytic in
the closed unit disk, except for the point $-1$, we in subsection~\ref{rigorous analysis} prove
that it asymptotically reduces to a constant. This implies that the perturbation just leads to a shift in
space with respect to the unperturbed propagating circle. In subsection~\ref{numerical}, we present numerical solutions of the linear evolution equations. These calculations support the asymptotic results
derived in the previous subsections. For disturbances, not necessarily localized, we present evidence
that on any part of the interface not containing a neighborhood of the bubble rear, the decay rate of the
disturbance matches what is expected from the prior spectral analysis \cite{PartI}.

Section~\ref{Sec:Nonlin} presents a numerical study of the nonlinear evolution for different perturbations. We first consider perturbations of a circular bubble. It is shown that the circular bubble can be nonlinearly stable if the perturbation is small. However, when the perturbation is large enough, the front may start to branch. Furthermore, we study the nonlinear evolution for more general initial configurations. It is shown that the formation of a cusp precisely on the back side of the moving body can not be excluded. We also observe that the body might split into two parts.

\section{Equations resulting from conformal mapping} \label{equations}

As already explained repeatedly~\cite{meul04, meul05, eber07, PartI}, we assume the streamer to be a simply connected compact domain ${\cal D}$ in the $(x, y)$-plane. The area of ${\cal D}$ is conserved under the dynamics and equals $\pi$ in dimensionless units. Identifying the $(x, y)$-plane with the closed complex plane $z = x + iy$, we introduce a conformal map $f (\omega, t)$ that maps the unit disk ${\cal U}_\omega$ in the $\omega$-plane to the complement of ${\cal D}$ in the $z$-plane
\begin{equation} \label{Apatit}
z = f (\omega, t) = \frac {a_{- 1} (t)}{\omega} + \hat f (\omega, t), \quad a_{- 1} (t) > 0 \,.
\end{equation}
The Laplace equation (\ref{Amazonit}) and the boundary condition (\ref{Antimonit}) are incorporated in the
definition of a complex potential $\Phi (\omega, t)$.
\begin{equation} \label {Aquamarin}
\Phi (\omega, t) = \frac{a_{- 1} (t)}{\omega} + \hat \Phi (\omega, t) \,.
\end{equation}
Both functions $\hat f (\omega, t)$ and $\hat \Phi (\omega, t)$ are analytic for $\omega \in  {\cal
U}_{\omega}$. The physical potential $\varphi (z, t)$ is related to $\Phi (\omega, t)$ as
\begin{equation} \label{Aragonit}
{\rm Re} \ \Phi (\omega, t) = \varphi (f (\omega, t), t) \,.
\end{equation}
The remaining boundary conditions (\ref{Achat}), (\ref{Amethyst}) take the form
\begin{eqnarray}
\label{Girasol} {\rm Re} \left[ \frac{\partial_{t} f}{\omega \partial_{\omega} f} \right] \
         =&  {\rm Re} \left[
\begin{displaystyle}
         \frac{\omega \partial_{\omega} \Phi}{| \partial_{\omega}f |^2}
\end{displaystyle}
         \right],  &\omega \in \partial \,{\cal U}_{\omega} \, \\[2.5ex]
\label {Azurit} | \partial_{\omega} f | {\rm Re} \ \Phi \ = & - \epsilon\, {\rm Re} \left[ \omega
\partial_{\omega} \Phi \right],  &\omega \in \partial \,{\cal U}_{\omega} \,.
\end{eqnarray}
The problem reduces to solving these two equations, respecting the analyticity properties of $f$ and $\Phi$.

A simple solution corresponding to a steadily translating circle is given by
\begin{eqnarray} \label{Bergkristall}
\begin{array}{ccl}
f^{(0)} (\omega, t) &=&
\begin{displaystyle}
\frac{1}{\omega} + \frac {2 t}{1+ \epsilon}
\end{displaystyle}
\\[2.5ex]
\Phi^{(0)}  (\omega, t) &=&
\begin{displaystyle}
\frac {1}{\omega} - \frac {1 - \epsilon}{1 + \epsilon} \ \omega \,.
\end{displaystyle}
\end{array}
\end{eqnarray}
In physical space it describes a unit circle moving with constant velocity $2/(1 + \epsilon)$ in
$x$-direction. For small and smooth distortions of this circle, it is appropriate to look for solutions of
the form
\begin{eqnarray} \label{Bernstein}
\begin{array}{ccl}
f (\omega, t)         & = & f^{(0)} (\omega, t) + \eta \ \beta (\omega, t)
\\[1.5ex]
\Phi (\omega, t)   & = &  \Phi^{(0)} (\omega, t) + \eta \ \frac{2}{1 + \epsilon} \ \chi (\omega, t)  \,,
\end{array}
\end{eqnarray}
where $\beta (\omega, t)$ and $\chi (\omega, t)$ are analytic in ${\cal U}_{\omega}$ and $\eta$ is a small
parameter. Since the area is conserved, it can be shown that the residue $1$ of the pole in
(\ref{Bergkristall}) remains unchanged to first order in $\eta$. Substituting (\ref{Bernstein}) into
equations (\ref{Girasol}), (\ref{Azurit}) we in first order in $\eta$ find a system of two partial
differential equations, from which $\chi$ can be eliminated. The final equation for $\beta$ takes the form
\begin{equation} \label {Beryll}
{\cal L}_\epsilon \ \beta = 0
\end{equation}
\begin{equation} \label{Biotit}
{\cal L}_\epsilon = \frac {\epsilon}{2}\  \partial_\omega \  (\omega^2 - 1)\  \omega \, \partial_\omega +
\epsilon \, \partial_\omega \,\omega \,  \partial_\tau + \partial_\tau - \partial_\omega \,,
\end{equation}
where we introduced the rescaled time variable
\begin{eqnarray} \label {Blauquarz}
\tau =  \frac {2}{1 + \epsilon} \ t \,.
\end{eqnarray}
Eqs.~(\ref {Beryll}), (\ref {Biotit}) determine the linearized evolution that will be discussed in
Section~\ref{analysis}. We will assume that the initial interface is analytic, i.e., that all singularities
of $\beta (\omega, 0)$  are outside the closed unit disk $\overline{\cal U}_{\omega}$, though much of the
analysis is valid for a sufficiently smooth interface as well\footnote{Analyticity is not crucial, except in
\S \ref{smoothness}, \ref{rigorous analysis}.}.

\section {Analysis of infinitesimal perturbations} \label{analysis}

\subsection{Summary of previous results} \label{previous results}

In part I~\cite{PartI} we have analyzed the eigenvalue problem, resulting from Eqs.~(\ref {Beryll}), (\ref
{Biotit}) via the ansatz $\beta (\omega, \tau) = e^{\lambda \tau} \beta_{\lambda} (\omega)$. We have shown
that the spectrum is purely discrete and that the real part of all eigenvalues $\lambda_{n}$ is negative,
except for the trivial value $\lambda_{0} = 0$, which corresponds to a simple shift of the circle. An infinite set of real negative eigenvalues was found. All
eigenfunctions, except for $\beta_{\lambda _0} (\omega) = {\makebox {const}}$, are singular at $\omega = -1$
at the back of the circle. Thus the expansion of a regular initial condition in terms of eigenfunctions has
to break down in the neighborhood of $\omega=-1$, which indicates that in that neighborhood some anomalous
relaxation shows up. Furthermore, we found that as $\epsilon \downarrow 0$, any eigenvalue $\lambda_{n}$
tends to zero and the corresponding eigenvector $\beta_{\lambda} (\omega)$ tends to a constant. A similar behavior of the spectrum was found for a steadily moving circle in a Hele-Shaw cell with surface tension regularization
\cite{TanveerSaffman} and this degeneracy is not unexpected since the unregularized problem $(\epsilon=0)$
is mathematically ill-posed \cite{Howison, FokasTanveer}.

Here, we consider the initial value problem defined by Eqs.~(\ref {Beryll}), (\ref {Biotit}). Our analysis
is guided by previous results \cite{meul05, eber07} on the special case $\epsilon = 1$ where the general
time dependent solution is known analytically; it is
\begin{equation} \label{Calcit}
\beta (\omega, \tau) = \frac {1}{\omega^2} \int \limits_{0}^{\omega}  \omega\,' \, G \left(
                         \frac{\omega\,' + T}{1 + \omega\,' \, T} \right)
                       d \omega\,' \,,
\end{equation}
where the function $G (\omega)$ is given by the initial condition,
\begin{equation} \label{Charoit}
G (\omega) = (2 + \omega \ \partial_{\omega}) \beta (\omega, 0) \,,
\end{equation}
and $T (\tau)$ is defined as
\begin{equation} \label{Nigrin}
T (\tau) = \tanh \frac{\tau}{2} \,.
\end{equation}
The properties of these solutions are discussed and visualized in detail in~\cite{meul05,eber07}. Here we in
particular note that the essential time dependence of $\beta (\omega, \tau)$ is contained in the
transformation
\begin{equation} \label{Citrin}
\zeta = \frac{\omega + T (\tau)}{1 + \omega T (\tau)} \,.
\end{equation}
$\zeta (\omega, T)$, $0 \le T \le 1$, defines a one-parameter family of automorphisms of the unit disk, with
fixed points $\omega = \pm 1$. The point $\omega = 1$ is stable, whereas $\omega = - 1$ is unstable in the
following sense: as $\tau \to \infty$, i.e. $T \to 1$, all the complex $\omega$-plane, except for $\omega =
- 1$, is mapped into a neighborhood of $\zeta = + 1$. This results in an advective dynamics. Any
perturbation not centered precisely at $\omega = 1$ is advected towards $\omega = -1$, where it vanishes
asymptotically. As $\tau \to \infty$, only a shift of the circle is left:
\begin{equation} \label{Florstein}
\lim \limits_{\tau \to \infty} \beta (\omega, \tau) = \frac {G(1)}{2} \,.
\end{equation}
However, it is to be noted that the limit is not uniform, and no matter how large $\tau$ is, there is a
neighborhood of $\omega=-1$, where $\beta (\omega, \tau)$ may change dramatically. We note that advection of
distortions from the front to the sides has been observed in viscous fingering and crystal growth models
with surface tension and has been derived from somewhat heuristically simplified models
\cite{Bensimon_etal, DeGreg86}. We further note that in the limit $\epsilon \to \infty$ a purely advective dynamics
results \cite{eber07}:
\begin{displaymath}
\beta (\omega, \tau) = \tilde \beta (\zeta (\omega, T (\tau)), \ \epsilon = \infty \,.
\end{displaymath}
Expecting the automorphism $\zeta (\omega, T)$ and the resulting advective dynamics to play an important
role also for $\epsilon \neq 1$ we transform the PDE (\ref{Beryll}), (\ref {Biotit}) from variables
$(\omega, \tau)$ to variables $(\zeta, T)$, introducing the notation
\begin{equation} \label{Diamant}
\beta (\omega, \tau) = \tilde \beta (\zeta (\omega, T (\tau)), T (\tau)) \,.
\end{equation}
This results in the normal form of a hyperbolic PDE:
\begin{equation} \label{Dioptas}
\left\{ \epsilon h (\zeta, T) \partial_{T} \ \partial_{\zeta} + \frac {\partial h (\zeta T)}{\partial T} \
    \partial_{\zeta} + (1 + \epsilon) \partial_{T} \right\} \tilde \beta (\zeta, T) = 0 \,,
\end{equation}
where
\begin {equation} \label {Disthen}
h (\zeta, T)  =  \frac {(\zeta - T) ( 1 - T \zeta)}{1 - T^2} =
     \omega (\zeta, T) \left [ \partial_{\zeta} \ \omega (\zeta, T) \right]^{- 1} \,.
\end{equation}

\subsection{Localized perturbations; rigorous results} \label{stronglyrigor}

Consider for general $\epsilon>0$ an initial perturbation that is centered at $\zeta=\zeta_c= e^{i \psi_c}$
for $\psi_c \ne 0$ and has `width' $\gamma$ in the sense that ${\tilde \beta} ( e^{i \psi_c + i \gamma
\chi}, 0) $ decays rapidly with $ |\chi|$ when $|\chi| \gg 1$. The decay rate will be specified more
precisely below Eq.~(\ref{eqGasymp}). To study this problem, we first write (\ref{Dioptas}) as an
integral equation:
\begin{equation}
\label{int.1} {\tilde \beta}_\zeta (\zeta, T) = \zeta^{1/\epsilon} h^{-1/\epsilon} (\zeta, T) {\tilde
\beta}_\zeta (\zeta, 0) - \frac{(1+\epsilon)}{\epsilon h^{1/\epsilon} (\zeta, T) } \int_0^T {\tilde \beta}_s
(\zeta, s) h^{-1+1/\epsilon} (\zeta, s) ds,
\end{equation}
where ${\tilde \beta}_\zeta$ and ${\tilde \beta}_s$ denote derivatives of ${\tilde \beta}$. Then integration
by parts in s replaces ${\tilde \beta}_s$ by $\tilde \beta$ which is written as
$$ {\tilde \beta}( \zeta, T) = \int_{\zeta_0}^\zeta {\tilde \beta}_\zeta
(\zeta', T) d\zeta' ~+~ {\tilde \beta} (\zeta_0, T) .$$ Here $\zeta_0 = e^{i \psi_0}$ is a reference point
in the tail of the perturbation chosen such that $0< \psi_0 < \psi_c \le \pi$. We assume that
$\frac{\psi_c-\psi_0}{\gamma} $ is so large that ${\tilde \beta} (\zeta_0, 0)$ is negligible.

Then, after some algebraic manipulation, we are able to rewrite (\ref{int.1}) as the following equation for
\begin{equation}
\label{eqG} {\hat G}(\chi, T) \equiv {\tilde \beta}_\zeta \left ( e^{i \psi_c + i \gamma \chi} , T \right ).
\end{equation}
\begin{eqnarray}
\nonumber {\hat G}(\chi, T) \ = \ {\hat G}^{(0)} (\chi, T) + \int_0^T
\int_{-\frac{\psi_c-\psi_0}{\gamma}}^{\chi} \mathcal{K}_1 (\chi, \chi',
T, s) {\hat G}(\chi', s) d\chi' ds  \\
\ +\ \int_{-\frac{\psi_c-\psi_0}{\gamma}}^{\chi} \mathcal{K}_2 (\chi, \chi', T) {\hat G}(\chi', T) d\chi'
\equiv {\hat G}^{(0)} (\chi, T) +  \mathcal{L} \left [ {\hat G} \right ] (\chi, T), \label{int.2}
\end{eqnarray}
where, with the understanding that $\zeta=e^{i \psi_c + i \gamma \chi}$, $\zeta'=e^{i \psi_c + i \gamma
\chi'}$,
\begin{eqnarray}
\nonumber \mathcal{K}_1 (\chi, \chi', T, s) &=& \frac{i \gamma \zeta' (1-\epsilon^2) (1-T^2)^{1/\epsilon}}{
\epsilon^2 \left \{ (\zeta-s) (1-\zeta s)\right \}^{1/\epsilon}} \left ( \frac{(\zeta-s)(1-s
\zeta)}{(1-s^2)} \right )^{-1+1/\epsilon}
\\
&& \hspace{3cm}\times \left[ \frac{1}{s-\zeta} + \frac{\zeta}{s \zeta-1} - \frac{2 s}{s^2-1} \right],
\label{eqK1}
\\
\label{eqK2} \mathcal{K}_2 (\chi, \chi', T) &=& -i \gamma \zeta'\frac{(1+\epsilon) (1-T^2)}{ \epsilon
(\zeta-T) (1-T \zeta) } \,,
\end{eqnarray}
and
\begin{eqnarray}
\nonumber {\hat G}^{(0)} (\chi, T) &=& h^{-1/\epsilon} (\zeta, T) \zeta^{1/\epsilon} {\tilde \beta}_\zeta
(\zeta, 0) - \frac{(1+\epsilon) {\tilde \beta} (\zeta_0, T) }{\epsilon h(\zeta, T)} + \frac{(1+\epsilon)
{\tilde \beta} (\zeta, 0) \zeta^{1/\epsilon} }{
\epsilon \zeta h^{1/\epsilon} (\zeta, T)} \\
&+& \frac{(1-\epsilon^2)}{\epsilon^2 h^{1/\epsilon} (\zeta, T) } \int_0^T h^{-2+1/\epsilon} (\zeta, s) ~h_T
(\zeta, s) ~{\tilde \beta} (\zeta_0 , s) ds \,. \label{eqG0}
\end{eqnarray}
With ${\hat G}^{(0)} (\chi, T)$ considered known\footnote{Since ${\tilde \beta} (\zeta_0, T)$ cannot be
determined without considering the full non-local problem on $|\zeta|=1$, part of the expression
(\ref{eqG0}) for ${\hat G}^{(0)}$ is not known. Nonetheless, if a disturbance is localized, the contribution
to ${\hat G}^{(0)}$ from ${\tilde \beta} (\zeta_0, T)$ will be relatively small. In any case, in order to
study the evolution in the $\chi$-scale, we are not prevented from considering ${\hat G}^{(0)}$ as known.}
we determine the solution ${\hat G} (\chi, T)$ to the integral equation (\ref{int.2}) for $\chi \in \left [
- \frac{\psi_c-\psi_0}{\gamma}, \chi_R \right ]$, $T \in [0, T_{0}]$, where $\chi_R$ and $T_{0} < 1$ are
some suitably chosen positive values independent of $\gamma$. Now it is clear from the expression for
$\mathcal{K}_1$ and $\mathcal{K}_2$ that they are uniformly small in the $\| . \|_\infty$ norm when
$\frac{\gamma}{\epsilon^2}$ is sufficiently small. We now choose the norm
\begin{equation}
\label{eqGasymp} \| {\hat G} \| \equiv \sup_{T \in [0, T_0]} \sup_{\chi \in \left
[-\frac{\psi_c-\psi_0}{\gamma}, \chi_R \right ]} W (\chi) | {\hat G} (\chi, T)|~~,
\end{equation}
where the positive weight function $W (\chi)$ obeys
\begin{displaymath}
W(\chi) \int_{-\infty}^{\chi_R} W^{-1} (\chi') d \chi' < C < \infty
\qquad\chi\le\chi_R.
\end{displaymath}
For example, $W(\chi) = e^{ -\chi}$ for $\chi \le 0$ and 1 for $\chi >0$ would suffice for our analysis. We
define ${\hat G}$ to be localized if $\|{\hat G} \|$ is finite, and $\zeta_0$, $T_0$ can be chosen such that $\tilde\beta(\zeta_0,T)$ is negligibly small for $T\in[0,T_0]$. Now it is clear from (\ref{int.2}) that the linear operator $\mathcal{L}$ has the contractive property
\begin{equation}
\| \mathcal{L} [{\hat G}_1-{\hat G}_2] \| \le C \frac{\gamma}{\epsilon^2} \| {\hat G}_1 - {\hat G}_2 \|.
\end{equation}
It follows that there exists a unique solution to the integral equation (\ref{int.2}) if $\gamma/\epsilon^2$
is small enough and that for $\gamma/\epsilon^2 \ll 1$
\begin{displaymath}
{\hat G}(\chi, T) \sim {\hat G}^{(0)} (\chi, T) \,,
\end{displaymath}
provided $\chi$ and $T$ are in the above specified range. For a perturbation localized in the sense given above, our result reduces to
\begin{equation}
\label{Johnit} {\hat G} (\chi, T) \sim h^{-1/\epsilon} \left( e^{i \, \psi_0}, T \right) \tilde
\beta_{\zeta}
     \left( e^{i \, (\psi_0 + \gamma \, \chi)}, 0 \right) \,.
\end{equation}
We note that ${\mathcal L} [{\hat G}] (\chi, T)$ in general will not vanish for $\chi \to \infty$. This is
the reason for restricting $\chi$ to the interval given above and indicates that for $T > 0$ the localized
perturbation will sit on top of a dynamically generated delocalized background of amplitude $\sim
\gamma/\epsilon^2$.

A detailed discussion of the result (\ref{Johnit}) will be presented in the next subsection.

\subsection{Localized perturbations; formal intuitive arguments}\label{strongly}

It is useful to obtain the result (\ref{Johnit}) through a more formal, yet intuitive, reasoning. This will
also be helpful in our subsequent treatment of the long-time asymptotics in the anomalous region near the
back of the bubble. We again restrict the analysis to the unit circle $\omega = e^{i \alpha}, \alpha \in
\mathbb{R}$, or correspondingly to $\zeta = e^{i \psi}, \psi \in \mathbb{R}$. According to Eq.~(\ref
{Citrin}), the two angular coordinates $\alpha$ and $\psi$ are related through
\begin{equation} \label{Epidot}
\alpha = \arctan \frac {\left ( 1 - T^2 \right) \sin \psi}{\left( 1 + T^2 \right) \cos \psi - 2 T} \,.
\end{equation}
Initially, (at $T = 0$), $\alpha$ and $\psi$ obviously are identical. In terms of $\psi$, the PDE
(\ref{Dioptas}) takes the form
\begin{equation} \label{Falkenauge}
\left \{ \epsilon \ \hat h (\psi, T) \partial_{T} \ \partial_{\psi} + \frac {\partial \ \hat h (\psi, T)}{
\partial T} \
    \partial_{\psi} + i (1 + \epsilon)\ \partial_{T} \right \} \tilde \beta \left( e^{i \psi}, T \right)
    = 0 \,,
\end{equation}
where
\begin{equation} \label{Zirkonia}
\hat h (\psi, T) = \left( \partial_{\psi} \ \alpha \right)^{- 1} =  \frac{(1 - T)^2 + 4 T \sin^2 \psi/2} {1
- T^2} \,.
\end{equation}
We now search for a solution that during its evolution stays localized near a fixed angle $\psi_{c}$, with
an angular width $\gamma\ll\pi$. We use the ansatz
\begin{equation} \label{Fluorit}
\tilde \beta \left( e^{i \psi}, T \right) =  \tilde \beta_{\scriptstyle {loc}} (\chi, T) \,,
\end{equation}
where again
\begin{equation} \label{Granat}
\chi = \frac {\psi - \psi_{c}}{\gamma}\,,
\end{equation}
and $\tilde \beta_{\scriptstyle{loc}} (\chi, T)$ is assumed to vanish rapidly for $| \chi | > 1$. With this
ansatz, Eq.~(\ref{Falkenauge}) takes the form
\begin{eqnarray} \label{Heliotrop}
\left[ \epsilon \ \hat h (\psi_c + \gamma \chi, T)\ \partial_{T} \ \partial_{\chi} + \left(
   \partial_T \ \hat h (\psi_c + \gamma \chi, T) \right) \ \partial_{\chi} + i \gamma (1 + \epsilon)
   \ \partial_{T} \right] \nonumber \\
       \cdot \ \tilde \beta_{\scriptstyle{loc}} (\chi, T) = 0  \,.
\end{eqnarray}
For $\gamma \ll \pi$ we neglect the term $i \gamma (1+\epsilon) \partial_T$ and the $\chi$-dependence in the
argument of $\hat h$ to find an approximate solution of the form
\begin{equation} \label{Holzstein}
\tilde \beta_{\scriptstyle{loc}} (\chi, T) = \hat h^{-1/\epsilon} (\psi_{c}, T) \ \tilde
\beta_{\scriptstyle{loc}} (\chi, 0),
\end{equation}
which is the same as (\ref{Johnit}).

Before we evaluate this result we briefly discuss its limitations, as resulting from the present derivation.
In view of the assumptions $\gamma \ll \pi,$ and $| \chi | \lesssim 1$, the use of the zero order result
$\hat h (\psi)\approx \hat h (\psi_{c})$ is justified provided
\begin{eqnarray} \label{Howlith}
(1-T)^2 + 4 T \sin^2 \frac{\psi_{c}}{2} \gg 2 T \gamma \chi \sin \psi_{c} + T (\gamma \chi)^2 \cos \psi_{c}
\, .
\end{eqnarray}
This is valid for all times provided $|\psi_{c}| \gg \gamma$, i.e., for initial conditions $\tilde
\beta_{\scriptstyle{loc}} (\chi, 0)$ which essentially vanish in the forward direction $\psi = 0$. For
$\psi_{c}\approx 0$ the condition (\ref{Howlith}) is violated if $(1-T)$ becomes of the order $\gamma$, and
therefore the approximation becomes invalid in the large-time limit $T (\tau) \to 1$. This special role of
perturbations in the forward direction is not unexpected since for such perturbations advection is
ineffective.

Neglecting the term $\sim i \gamma (1 + \epsilon) \partial_{T}$ has more serious consequences. Substituting
into Eq.~(\ref{Heliotrop}) an ansatz of the form
\[ \tilde \beta_{\scriptstyle {loc}} (\chi, T) = \tilde \beta^{(0)} (\chi, T) + \gamma \ \tilde \beta^{(1)}
(\chi, T)
              + {\cal O} (\gamma^{2})\]
one finds that the result for $\tilde\beta^{(1)}$ violates the condition $\tilde \beta^{(1)} (\chi, T)
\approx 0$ for $| \chi| \gg 1$. A localized initial condition dynamically generates a delocalized
contribution, with an amplitude proportional to $\gamma/\epsilon^2$, in full accord with the rigorous
discussion of the previous subsection. Again this result is not unexpected since the eigenfunctions of the
operator ${\cal L}_{\epsilon}$, Eq.~(\ref {Biotit}), are delocalized. Assuming that we can expand an
initially localized perturbation in terms of eigenfunctions we must expect that the balance of the expansion
coefficients $a_n \ e^{\lambda_n\, \tau}$, which for $\tau=0$ leads to localization, is destroyed by the
time evolution. With these limitations in mind, we now discuss the result (\ref{Holzstein}).

According to Eq.~(\ref{Holzstein}), if expressed in the variable $\zeta = \exp \, (i (\psi_{c} + \gamma
\chi))$ the evolution of the perturbation is most simple. Neither its position $\psi_{c}$ nor its shape
$\hat \beta_{\scriptstyle{loc}} (\chi, 0)$ change. Only the overall amplitude $\hat h ^{-1/\epsilon}$ varies
with time. For $0 < | \psi_{c} | < \pi/2$, i.e., if $\psi_{c}$ is at the front half of the circle, $\hat
h^{-1/\epsilon}$ increases up to a time $\tau_m$ given by
\begin{equation} \label{Zinnstein}
\left( T^2 (\tau_m) + 1\right) \cos \ \psi_{c} - 2 T (\tau_m) = 0 \,,
\end{equation}
and then decreases again. For $| \psi_{c} | > \pi/2,  \hat h^{-1/\epsilon}$ decreases monotonically. For any
$\psi_{c} \ne 0$, we find the asymptotic behavior
\begin{equation} \label{Jade}
\hat h^{-1/\epsilon} \left ( \psi_c, T (\tau) \right ) \sim \frac {e^{- \tau/\epsilon}}{\sin^2  \psi_c /
2}\, \quad {\makebox{for}} \quad \tau \to \infty \, .
\end{equation}
For a perturbation centered precisely at the back of the circle $(\psi_{c} = \pi)$, exponential relaxation
\begin{equation} \label{Jaspis}
\hat h^{- 1/\epsilon} \left (\pi, T (\tau) \right) = \left ( \frac {1-T}{1+T} \right )^{1/\epsilon} =
e^{-\tau/\epsilon}
\end{equation}
holds for all $\tau$. We recall that the localized approximation must break down it $\hat h^{-1/\epsilon}$
becomes of the order of the amplitude of the delocalized background. Nevertheless we will argue in
subsection\,\ref{asymptotic} that a contribution with asymptotic time behavior $e^{-\tau/\epsilon}$
generally shows up.

Using Eq.~(\ref{Epidot}) to transform back to $\omega = e^{i \alpha}$ we see that the center $\alpha_c (T
(\tau))$ is convected along the circle, reaching $\pm \,\pi$ for $\tau \to \infty$. A little calculation
yields the velocity of this advection
\begin{equation} \label{Karneol}
\frac{d}{d\tau} \alpha_{c} \left ( T (\tau) \right ) = \sin \alpha_{c} \left (T (\tau) \right)\, .
\end{equation}
This result has a simple interpretation. Recalling that we are working in a frame moving with the velocity
${\bf v} = \hat{\bf x}$ of the unperturbed circle, we identify the velocity (\ref{Karneol}) as the
projection of ${\bf v}$ onto the tangent to the circle at the instantaneous location of the perturbation.

In terms of $\alpha_{c}$ the overall amplitude of the perturbation takes the simple form
\begin{equation} \label{Katzenauge}
\hat {h}^{-1/\epsilon} = \left ( \frac{\sin \alpha_c \left ( T (\tau) \right )}{\sin \alpha_{c} (0)} \right
)^{1/\epsilon} \,.
\end{equation}
It increases as long as the perturbation is on the front half of the circle and decreases on the backside.
The maximum, reached for $\alpha_{c} \left ( T (\tau) \right ) = \pm \,\pi/2$, strongly depends on the
initial position $\alpha_{c} (0) \equiv \psi_{c}$.

Defining the scale factor of the width of the perturbation as
\begin{equation} \label{Kunzit}
\Gamma = \frac{\partial \alpha}{\partial \psi} \Bigr |_{\psi = \psi_{c}}
\end{equation}
we find
\begin{equation} \label{Labradorit}
\Gamma = \hat {h}^{-1} \left ( \psi_{c}, T (\tau) \right) = \frac{\sin \alpha_c \left( T (\tau)
\right)}{\sin \alpha_c (0)} \,.
\end{equation}
Thus the width behaves similarly to the amplitude, except that for $\epsilon \ll 1$ it varies much less. For
$\tau \to \infty$ it vanishes like $e^{-\tau}$.

So far we considered perturbations localized away from the tip $\psi_{c} = 0 = \alpha_{c} (0)$ of the
circle. For $\psi_{c} = 0$, Eq.~(\ref{Holzstein}) still holds for times such that
\begin{displaymath}
1 - T (\tau) \gg \gamma \, ,
\end{displaymath}
cf.~Eq.~(\ref{Howlith}). It describes the initial increase and broadening of the perturbation. Advection, of
course, is absent. For $1 - T \approx \gamma$ the width becomes of order $1$ and the local approximation
clearly becomes invalid.

On the qualitative level these results are most similar to the exact results found for $\epsilon = 1$
\cite{meul05, eber07} and resemble the dynamics of a localized perturbation found in the context of viscous
fingering \cite{DeGreg86}.

\begin{figure}[t]
\begin{center}
\includegraphics [scale=0.65]  {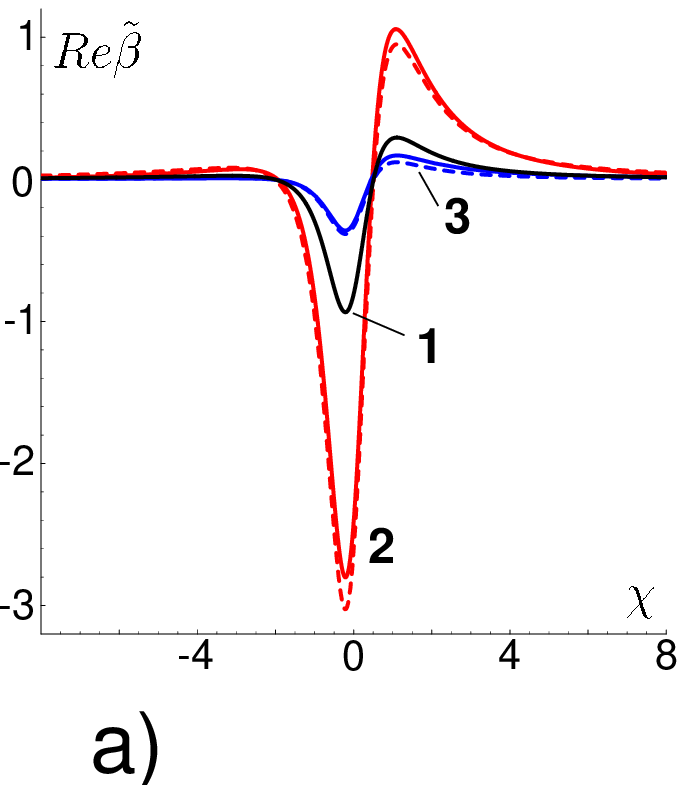} \qquad  \qquad
\includegraphics [scale=0.9]  {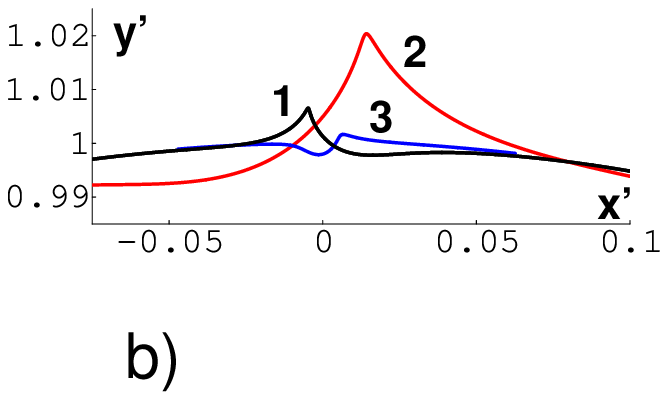}
\caption{Evolution of a strongly localized perturbation for $\epsilon = 1$. Curves 1, 2, 3 correspond to
times $\tau = 0$, 1.84 and 4.59, respectively. a) Re~$\tilde \beta$ as function of the comoving angular
coordinate $\chi=(\psi - \psi_c)/\gamma$. Broken lines: local approximation. Full lines: exact result with
the shift of the circle subtracted. b) Perturbed interface in the physical plane in the system of local
tangential and normal coordinates as explained in the text.} \label{fig1}
\end{center}
\end{figure}

The quantitative performance of the local approximation is illustrated in Fig.~\ref{fig1}, where for
$\epsilon = 1$ the exact evolution of a localized perturbation is compared to our approximation. From the
exact result (\ref{Calcit}) the contribution $G (T)/2$ representing a simple shift of the circle, has been
subtracted. The initial condition is chosen as
\begin{displaymath}
\tilde \beta (e^{i \psi}, 0) = \frac {\gamma^2}{ \bigl( e^{i \psi} - (1 + \gamma) \ e^{i \psi_c} \bigr)^2}
\end{displaymath}
with
\begin{displaymath}
\psi_{c} = - \frac {\pi}{10} \,, \quad \gamma = \frac{1}{200} \,.
\end{displaymath}
Fig.~\ref{fig1}a shows ${\rm Re}\, \tilde \beta$ as function of $\chi = (\psi - \psi_{c}) / \gamma$ for
three different times. Curve 1 shows the initial condition, where by construction the exact form and the
approximation coincide. Curve 2 shows the perturbation when it is largest, in $\omega$-space being located
near $\omega = - i$. Curve 3 is taken at some later time. Evidently in this example the local approximation,
(broken lines), is quite accurate. Very similar results are found for ${\makebox{Im}}\, \tilde \beta$, which
therefore is not shown. Fig.~\ref{fig1}b shows the effect of this perturbation in physical space. In
evaluating $z = f (\omega, t)$, Eq.~(\ref{Bernstein}), we choose the amplitude $\eta = 0.007 e^{-i
\psi_{c}}$. To combine the three curves into one plot, we introduced a time-dependent rotation of the
coordinate system such that $y'$ or $x'$ are measured along the normal or the tangent to the unperturbed
circle at the center of the perturbation, (i.e., at angle $-\alpha_{c} \,(T (\tau))$, since the inversion
contained in the conformal map induces a sign change of the angles). In this representation the exact
solution and the approximation cannot be distinguished within the resolution of the plot. We note that in
physical space the shape of the perturbation varies due to interference with the unperturbed circle.

\subsection{Asymptotic relaxation near $\omega = - 1$} \label{asymptotic}

In discussing the asymptotic relaxation we prefer to rewrite (\ref{Dioptas}) in terms of $\tau$, using $(1 -
T^2) \partial_{T} = 2 \partial_{\tau}$. Inserting the explicit form (\ref{Disthen}) of $h (\zeta, T)$ and
multiplying by $(1 - T^2)^2 / 2$ we find
\begin{eqnarray} \label{Lapislazuli}
\Bigl [ \epsilon (\zeta - T) (1-T\zeta) \partial_{\tau} \ \partial_{\zeta} +
\Big( 2 T \zeta- \frac{1}{2} (1+T^{2}) (1+\zeta^{2}) \Big) \partial_{\zeta} \nonumber \\
+\, (1+\epsilon) (1-T^{2}) \partial_{\tau} \Bigr ] \, \tilde \beta (\zeta, T) = 0 \, .
\end{eqnarray}
Here $T$ stands for
\begin{displaymath}
T = T (\tau) = \tanh \tau/2 = 1 - 2 e^{-\tau} + {\cal O} \left (e^{-2 \tau} \right ) \, ,
\end{displaymath}
cf.~Eq.~(\ref{Nigrin}). Keeping only the leading $\tau$-dependence in the coefficients of the derivatives,
we reduce Eq.~(\ref{Lapislazuli}) to
\begin{eqnarray} \label {Larimar}
\Big  [ - (\epsilon \partial_{\tau} + 1) ( 1-\zeta)^2 \ \partial_\zeta + 4 \, (1 + \epsilon) \ e^{-\tau}
\partial_{\tau}\Big] \tilde \beta (\zeta, T (\tau)) = 0 \,.
\end{eqnarray}
For $\tau \gg 1$ we neglect the term $e^{- \tau} \partial_{\tau} \ \tilde \beta$ to find
\begin{equation} \label{Malachit}
\tilde \beta (\zeta, T (\tau)) \sim e^{-\tau/\epsilon} \hat \beta_{0} (\zeta)+ \gamma_{0} \, ,
\end{equation}
where $\hat \beta_{0} (\zeta)$ and $\gamma_{0}$ depend on the initial condition $\tilde \beta (\zeta, 0)$
and of course cannot be fixed by this asymptotic argument.

Since the derivative $\partial_{\zeta}$ in Eq.~(\ref{Larimar}) is multiplied by $(1 - \zeta)^2$, the neglect
of the term involving $e^{- \tau} \partial_{\tau}$ can be justified only for $\zeta \not= + 1$. In terms of
\begin{equation} \label{Moldarit}
\omega = \frac {\zeta - T}{1 - T \zeta} = - 1 + 2 \ \frac {1 + \zeta}{1 - \zeta} \ e^{-\tau} + {\cal O}
\left (e^{-2 \tau} \right )
\end{equation}
this implies that we deal with a neighborhood of $\omega = -1$ that is contracted to this point like $e^{-
\tau}$. This range of $\omega$ is complementary to the region where an expansion in terms of eigenfunctions
can be expected to be valid asymptotically.

In the result (\ref{Malachit}) the $\zeta$-dependence is suppressed by a factor $e^{- \tau/\epsilon}$, which
for $\epsilon \ll 1$, $\tau \to \infty$, vanishes much faster than $e^{- \tau}$. Thus $\zeta$-dependent
corrections of order $e^{- \tau}$ will dominate the asymptotic relaxation at the back of the circle. Noting
the presence of $e^{-\tau}$ in the coefficients of the differential equation, it is natural to determine the
structure of these terms with the ansatz
\begin{equation} \label{Mondstein}
\tilde \beta \big( \zeta, T (\tau) \big) = e^{- \tau/\epsilon} \hat \beta (\zeta, \tau) + \hat \gamma
(\zeta, \tau),
\end{equation}
where
\begin{equation} \label{Mookait}
\hat \gamma (\zeta, \tau) = \sum \limits_{k=0}^{\infty} \hat \gamma_{k} (\zeta) \ e^{-k\, \tau} \,.
\end{equation}
From Eq.~(\ref{Lapislazuli}) with $\tilde \beta (\zeta, T (\tau))$ replaced  by $\hat\gamma (\zeta, \tau)$
we find
\begin{eqnarray} \label{Morganit}
&&\hat \gamma (\zeta, \tau) = \hat c_{0} + \hat c_{1} \ e^{- \tau} +
  \left( \hat c_{2} - \frac{1 + \epsilon}{1 - 2 \epsilon} \ \frac{4 \hat c_{1}}{1 - \zeta} \right) \ e^{- 2 \tau}
  \\&&\quad+ \left( \hat c_{3} - \frac{1 + \epsilon}{1 - 3 \epsilon} \ \frac {8 \hat c_{2}}{1 - \zeta} +
  \frac{(1 + \epsilon)^2}{(1 - 2 \epsilon) (1 - 3 \epsilon)} \ \frac{16 \hat c_{1}}{(1 - \zeta)^2} \right) \
  e^{- 3 \tau} + {\cal O} \Big( e^{-4 \tau} \Big) \,,  \nonumber
\end{eqnarray}
where the $\hat c_{k}$ are integration constants. Generally $\hat \gamma_{k} (\zeta)$ is found to be a
polynomial in $(1 - \zeta)^{- 1}$ of degree $k - 1$. In this analysis we assumed $\epsilon \not= 1/n$. For
$\epsilon = 1/n$ the ansatz (\ref{Mookait}) has to be modified. In particular a term proportional to $\tau
e^{-n \, \tau}$ has to be included. We note that the exact result for $\epsilon = 1$ shows such a
contribution \cite{eber07}.

To transform our result back to $\omega$-space we introduce
\begin{eqnarray}
\label{Nephrit}
\delta \omega &=& (1 + \omega) \ e^\tau \,, \\[1.5ex]
\label{Obsidian} \gamma (\delta \omega, \tau) &=& \hat \gamma  \Big(\zeta (\delta \omega), \tau\Big) - \hat
\gamma ( - 1, \tau)\,.
\end{eqnarray}
Eq.~(\ref{Citrin}) yields
\begin{equation}\label{Onyx}
\frac{1}{1 - \zeta} = \frac {1 + \frac{T}{1 + T} \ \delta \omega}{2 - \delta \omega \, e^{- \tau}} =
\frac{1}{2} + \frac{1}{4} \ \delta \omega + {\cal O} (e^{- \tau}) \,.
\end{equation}
Thus $\gamma (\delta \omega, \tau)$ has an expansion of the form
\begin{equation} \label{Opal}
\gamma (\delta \omega, \tau) = e^{- \tau} \sum \limits_{k=1}^{\infty}
      c_k \big( \delta \omega \, e^{- \tau} \big)^k \ \big[1 + {\cal O} \big( e^{- \tau} \big) \big] \,,
\end{equation}
where the $c_k$ again depend on the initial condition. For $\epsilon \ll 1$ the terms of order $k <
1/\epsilon - 1$ dominate over the contribution $e^{- \tau/\epsilon} \hat \beta$. For $\epsilon < 1/2$ we
therefore  in a region of size $| 1 + \omega | = {\cal O} \ (e^{- \tau})$ near $\omega = - 1$ expect to see
a very smooth asymptotic relaxation of the interface, with only a few coefficients depending on the initial
condition. In contrast, for $\epsilon > 1/2$ the asymptotic relaxation is determined by the term $e^{-
\tau/\epsilon} \hat \beta_{0} (\zeta)$,  which will depend on the initial condition in a complicated way.
For $\epsilon = 1$ this is illustrated in Fig.~5.2 of Ref.~\cite{eber07}. In the next subsection we will
argue that the function $\hat \beta_{0} (\zeta)$ picks up contributions due to singularities of the initial
condition, which for $\tau \to \infty$ are driven towards $\omega = - 1$. We finally note that the results
discussed here resemble the behavior of the low order eigenfunctions $\beta_{\lambda} (\omega)$. As shown in
part I~\cite{PartI} of this series, these functions near $\omega = - 1$ develop a singularity of the form
$(1 + \omega)^{1/\epsilon + \lambda}$, implying that the derivatives at $\omega = - 1$ exist for all orders
$k < 1/\epsilon + \lambda$.

\begin{figure}[h]
\begin{center}
\includegraphics [scale=0.6]  {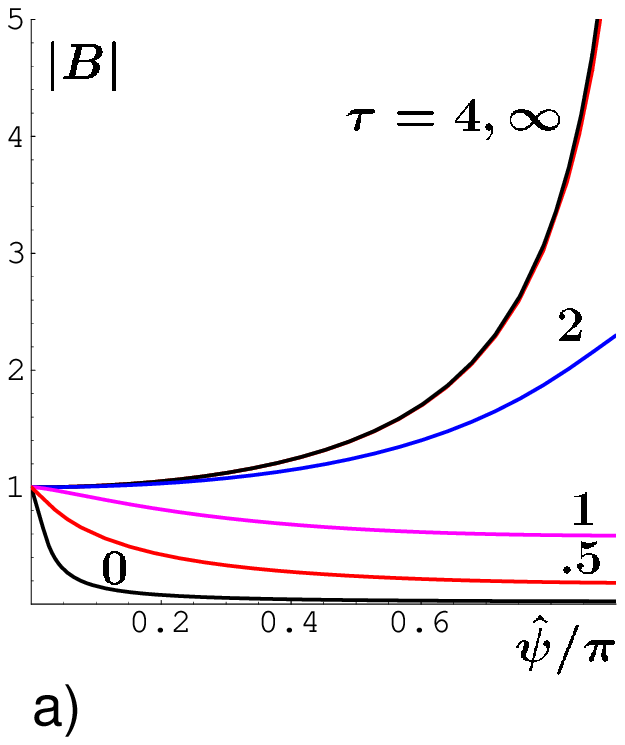}\quad
\includegraphics [scale=0.6]  {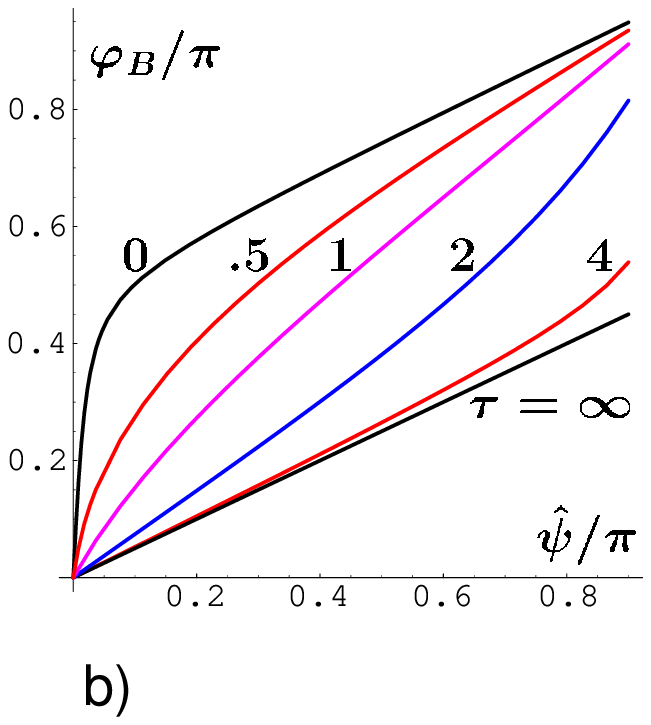}\quad
\includegraphics [scale=0.6]  {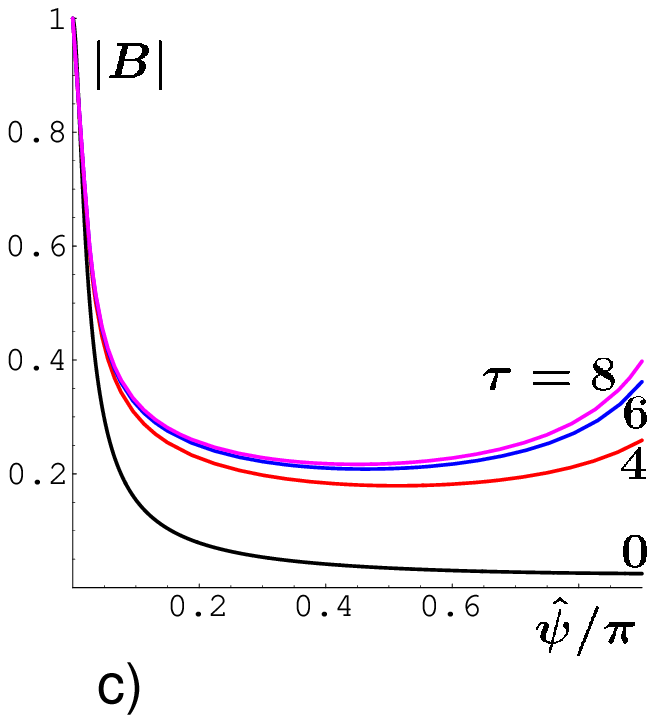}
\caption{Time evolution of the initial condition (\ref{Peridot}). a) $| B (\hat \psi, \tau) |$ as function
of $\hat \psi/ \pi$ for times $\tau$ as given and for $\epsilon = 0.1$. b) Phase of $B$ for the same values
of $\tau$ and $\epsilon$. c) Again $| B (\hat \psi, \tau) |$ as function of $\hat \psi/ \pi$, but now for
$\epsilon= 0.8$. \label{fig2}}
\end{center}
\end{figure}

To illustrate our results we consider a perturbation centered at $\omega = - 1$. As initial condition we
choose
\begin{equation} \label{Peridot}
\tilde \beta (\zeta, 0) = \frac{\gamma}{\gamma - \zeta} \quad, \quad \gamma = 1.05 \,,
\end{equation}
and we calculate the function
\begin{equation} \label{Pietersit}
B (\hat \psi, \tau) = \frac{\tilde \beta ( - e^{i \hat \psi}, T (\tau)) - \tilde \beta (0, T (\tau))}
                {\tilde \beta (- 1, T (\tau)) - \tilde \beta (0, T (\tau))} \,.
\end{equation}
We expect to find the limiting behavior
\begin{equation} \label{Prasem}
B (\hat \psi, \tau) \ {\footnotesize {\overrightarrow{\tau \to \infty}}} \ \frac {1}{1 + e^{i \hat \psi}} -
1 =
     \frac{e^{i \hat \psi/2}}{\cos \hat \psi/2} \,, \quad \epsilon < \frac{1}{2} \,,
\end{equation}
or
\begin{equation}\label{Pyrit}
B (\hat \psi, \tau)  \ {\footnotesize{\overrightarrow{\tau \to \infty}}} \ \frac{\hat \beta_0 (- e^{i \hat
\psi}) - \hat \beta_0 (0)}{\hat \beta_0 (- 1) - \hat \beta_0(0)}\,, \quad \epsilon > \frac{1}{2} \,,
\end{equation}
respectively. Whereas $\hat \beta_0 (\zeta)$ depends on the initial condition, the limit (\ref{Prasem}) is
universal. The results shown in Figure \ref{fig2} conform to these expectations.

Fig.~\ref{fig2}a shows $| B (\hat \psi, \tau) |$ for several values of $\tau$ and for $\epsilon = 0.1$. It
illustrates the approach to the limiting form $1 / \cos (\hat \psi/2)$, which within the accuracy of the
plot is in fact reached for $\tau \approx 4$. Fig.~\ref{fig2}b shows the corresponding phase of $B (\hat
\psi, \tau)$. Here the approach to the limit is slower, but is definitely visible. Fig.~\ref{fig2}c shows
results for $| B (\hat \psi, \tau)|$, $\epsilon = 0.8$. Here $ |B (\hat \psi,  \tau) |$ seems to approach a
limiting curve which clearly shows remainders of the initial peak. (We should note that $B (\hat \psi,
\tau)$ is symmetric: $B (-\hat \psi, \tau) = B^{*} (\hat \psi, \tau)$, and that the peak at $\hat \psi = 0$,
of course, is rounded, which however is not visible on the scale of the plot). We finally recall that the
$\hat\psi$-range shown here in terms of $\omega = e^{i \alpha}$ corresponds to a small region near $\alpha =
\pi$. Specifically for $\tau = 4$ it corresponds to $\pi  \le \alpha \le 1.08\ \pi$.

\begin{figure}[h]
\begin{center}
\includegraphics [scale=0.7] {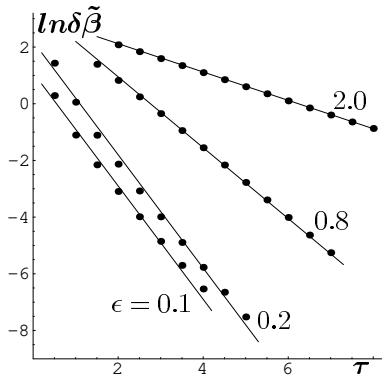}
\caption{$\ln \left[\tilde \beta \left(- 1, T (\tau) \right)- \tilde \beta \left( 0, T (\tau) \right)
\right]$ as a function of $\tau$ for several values of $\epsilon$. The initial condition is given in
(\ref{Peridot}). The lines indicate the expected slope $-2$ for $\epsilon < 1/2$ or $- 1/\epsilon$ for
$\epsilon > 1/2$, respectively. \label{fig3}}
\end{center}
\end{figure}

For the asymptotic relaxation our results predict
\begin{equation} \label{Rauchquarz}
\tilde \beta (\zeta, T (\tau)) \sim \left \{
\begin{array}{ll}
      e^{- 2 \tau} \,, & \epsilon < \frac{1}{2} \\ [2.0ex]
      e^{- \tau/\epsilon} \,, & \epsilon > \frac{1}{2} \\
     \end{array} \right.~~~\mbox{for }\tau\to\infty.
\end{equation}
This prediction is tested in Figure~\ref{fig3} by plotting results for $\ln [\tilde \beta (-1, T (\tau)) -
\tilde \beta (0, T (\tau)]$ as function of $\tau$ for several values of $\epsilon$. The expected behavior is
reasonably well observed.

\subsection{Analyticity of the interface} \label{smoothness}

If we assume the initial interface to be analytic, all singularities of the initial perturbation $\beta
(\omega, 0)$ have to be outside the closed unit disk, $\overline{\cal U}_{\omega}$. We here argue that under
the linearized dynamics the singularities stay outside $\overline{\cal U}_{\omega}$ for all finite times
$\tau$. For $\tau \to \infty$ they approach $\omega = - 1$, and contribute to the anomalous $e^{-
\tau/\epsilon} \ \hat \beta (\zeta)$ behavior found in subsection~\ref{asymptotic}.

This argument is based on the recurrence relation for the coefficients $b_{k} (\tau)$ in the Taylor expansion
\begin{equation} \label{Serpentin}
\beta (\omega, \tau) = \sum^{\infty}_{k=0} b_k (\tau) \omega^k \,.
\end{equation}
The evolution equation~(\ref{Beryll}), (\ref{Biotit}) yields
\begin{eqnarray} \label{Sugilith}
2 \partial_{\tau} b_{0} &=& {\frac{2+\epsilon} {1+\epsilon}} \ b_{1} \, ,
\\
2 \partial_{\tau} b_{k} &=& \frac {k+1}{1 + \epsilon + \epsilon k} \Bigl [ (2 + \epsilon + \epsilon k) \
b_{k + 1} - \epsilon (k- 1) b_{k-1} \Bigr ] \quad\mbox{for } k \ge 1 \,. \nonumber
\end{eqnarray}
The singularities of $\beta (\omega, \tau)$ are determined by the behavior of the $b_{k} (\tau)$ in the
limit $k \to \infty$. For simplicity, we consider an initial condition $\beta (\omega, 0)$ whose singularity
closest to $|\omega|=1$ is a branch point at $\omega_0$ with behavior $\beta (\omega, 0) \sim
{\rm const} (\omega_0 -\omega)^{\alpha-1}$ for nonintegral $\alpha$ or a pole with $\alpha$ a non-positive
integer. Then $b_{k} (0)$ for $k \gg1$ behaves as
\begin{displaymath}
b_k (0) = {\makebox{const}} \ \omega_0^{-k} k^{- \alpha} \left( 1 + {\mathcal O} \left( \frac{1}{k} \right)
\right) \,.
\end{displaymath}
We therefore make the ansatz
\begin{equation} \label{Magnesit}
b_{k} (\tau) = (-1)^{k} \ e^{- k f (\tau)} g (\tau) k^{- \alpha} \left [ 1 + \frac{1}{k} \Delta (\tau) +
{\cal O} \left( \frac{1}{k^2} \right) \right],
\end{equation}
where the factor $(- 1)^k$ is introduced since we expect the point $\omega = - 1$ to play a special role. We
will find that this ansatz is internally consistent provided $k \gg k_{0}$, where $k_{0}$ increases with
$\tau$. We shall conclude that for any $\tau > 0$,
\begin{equation} \label{Rhodonit}
| \, e^{- f (\tau)} \, | \, < 1 ,
\end{equation}
if this condition is satisfied initially. This implies that the singularity remains outside the unit disk
for all times and suggests that at least for initial conditions with a branch point, the interface will
remain analytic.

Substituting the ansatz (\ref{Magnesit}) into the recurrence relation (\ref{Sugilith}), we find
\begin{eqnarray} \label{Rubin}
&&\Bigl( - 2 \, \partial_\tau f + e^{-f} - e^f \Bigr) \left( 1 + \frac\Delta k \right) \nonumber \\
&&~~+ \frac 1 k \left[ 2 \, \partial_\tau\,  \ln \, g + \left(\frac 1 \epsilon + 1 - \alpha \right) \Bigl(
e^f + e^{-f} \Bigr) \right] = {\cal O} \left( \frac 1 {k^2} \right) .
\end{eqnarray}
The leading order yields
\begin{displaymath}
2 \, \partial_{\tau} f \, = \, e^{-f} - e^f \, ,
\end{displaymath}
with the solution
\begin{equation} \label{Saphir}
f (\tau) \, =\,  \ln \, \frac {1 - C e ^{-\tau}}{1 + C e^{-\tau}} \, ,
\end{equation}
where $C$ is some integration constant. $g (\tau)$ is determined by the next order:
\begin{equation} \label{Sardonyx}
g (\tau) \, = \, g(0) \, e^{\left(\alpha - 1- \frac{1}{\epsilon}\right) \tau} \Bigl( 1 - C^{2} \, e^{-2
\tau} \Bigr) ^{\alpha - 1 - \frac{1}{\epsilon}}  \, .
\end{equation}
Checking higher orders in an expansion in powers of $1/k$, one finds that neglecting such terms assumes $k
\gg \frac{e^{\tau}}{\epsilon} = k_{0}$. Combining our results we find the asymptotic behavior
\begin{equation} \label{Smaragd}
b_{k} (\tau) \sim (-1)^k \left( \frac{1 + C \, e^{-\tau}}{1 - C \, e^{-\tau}} \right)^k g(0) \,
e^{\left(\alpha - 1 - \frac1\epsilon\right) \tau} \left( 1 - C^2 \, e^{-2 \tau} \right) ^{\alpha - 1 - \frac
{1}{\epsilon}}\, k^{-\alpha}    \, .
\end{equation}
Regularity of the initial condition enforces
\begin{displaymath}
\left| \, e^{- f (0)} \, \right|  = \left| \, \frac {1 + C}{1 - C} \, \right| \,< 1 ,
\end{displaymath}
equivalent to Re$~C < 0$. With the form (\ref{Saphir}) of $f (\tau)$ this guarantees that condition
(\ref{Rhodonit}), $| e^{- f (\tau)} | < 1$, is fulfilled for all finite $\tau$. Thus for $\tau < \infty$ the
singularities of $\beta (\omega, \tau)$ stay at some finite distance from the unit disk and the interface
stays smooth. $f (\tau)$ vanishes for $\tau \to \infty$, indicating that a singularity reaches $\omega = -
1$.

In the above ansatz (\ref{Magnesit}), we assumed a particular type of branch point or a pole for $\beta
(\omega, 0)$ as the nearest singularity. Multiple singularities of this type can be accommodated in this
linear analysis using the superposition principle. Other singularities can be accommodated as well by
replacing $k^{-\alpha}$ by a more general $k$ dependence.

We now consider the limiting behavior of $b_{k} (\tau)$ for $\tau \to \infty$ more closely.
Eq.~(\ref{Smaragd}) yields
\begin{equation} \label{Sodalith}
b_{k} (\tau) \sim (- 1)^k \exp \left[ 2 \, C \, k \, e^{-\tau} \right] g(0)
    \left( k \, e^{- \tau} \right)^{1 + \frac {1}{\epsilon} - \alpha} k^{1 - \frac{1}{\epsilon}} \,.
\end{equation}
This result, however, for $\tau \to \infty$ is only valid for
\begin{equation} \label{Sonnenstein}
\eta = k \ e^{- \tau} \gg 1 \,,
\end{equation}
i.e., for extremely large $k$. To extend the analysis to values $\eta= k \ e^{- \tau} = {\cal O} (1)$ we
make the ansatz
\begin{equation} \label {Spinell}
b_k (\tau) \sim (- 1)^k \ k^{-\frac {1}{\epsilon} - 1} \ \tilde g (\eta, \tau) \,,
\end{equation}
which is motivated by Eq.~(\ref{Sodalith}). The recurrence relation (\ref{Sugilith}) takes the form
\begin{eqnarray*}
(2 \partial_{\tau} - 2 \eta \ \partial_{\eta})  \ \tilde g (\eta, \tau) & = &   \left( k + {\cal O} \left
(\frac{1}{k} \right ) \right) \ \tilde g \left(\eta - e^{- \tau}, \tau\right)
 \nonumber \\
& & -  \left( k + {\cal O} \left ( \frac{1}{k} \right ) \right) \ \tilde g \left(\eta + e^{- \tau},
\tau\right)
\nonumber \\
& = &   -  2 k \ e^{- \tau} \partial_{\eta} \ \tilde g (\eta, \tau) + {\cal O} \left ( \frac {1}{k}
\right )   \nonumber \\
& = &   -  2 \eta \ \partial_{\eta} \ \tilde g (\eta, \tau) + {\cal O} \left( \frac {1}{k} \right)  ,
\end{eqnarray*}
or
\begin{equation} \label {Thulit}
2 \ \partial_{\tau} \ \tilde g (\eta, \tau) = {\cal O} \left( \frac{1}{k} \right) ,
\end{equation}
equivalently. Thus to leading order in $1/k$, $\tilde g (\eta, \tau)$ is independent of $\tau$ and
Eq.~(\ref{Prasem}) reduces to
\begin{equation} \label {Tigereisen}
b_{k} (\tau) \sim (- 1)^k \, k^{ - \frac{1}{\epsilon} - 1}\, \tilde g_{0} \left( k \ e^{- \tau} \right) \,.
\end{equation}
Inspecting the terms of order $1/k$ one finds that this result asymptotically should be valid for $e^{-
\tau} \ll 1$ and $k \gtrsim e^{\tau} / \epsilon^2$.

The $b_{k} (\tau)$, Eq.~(\ref {Tigereisen}), can be interpreted as coefficients of a Taylor expansion with
respect to $\omega$ of the function $\hat \beta_{0} (\zeta (\omega, T))$ introduced in the previous
subsection, cf.~Eq.~(\ref{Malachit}). To show this we again introduce $\delta\omega$
\begin{displaymath}
\omega= - 1 + \delta \omega \ e^{- \tau}
\end{displaymath}
as defined in Eq.~(\ref{Nephrit}), and we approximately resum the Taylor expansion from $k = \eta_{0}
e^\tau$ to infinity, using the result (\ref{Tigereisen}).
\begin{eqnarray*}
&\sum \limits^{\infty}_{k=\eta_{0} e^\tau}  & b_{k}(\tau) \left (-1 + \delta \omega \ e^{-\tau}\right )^k \\
 & \approx & \int \limits^{\infty}_{\eta_{0} e^\tau}  d k \, k^{-\frac{1}{\epsilon}-1} \tilde g_{0}
     (k \ e^{-\tau}) \exp \left [ -k \ e^{-\tau} \delta \omega \right ] \\
& = & e^{-\tau/\epsilon} \int \limits^{\infty}_{\eta_{0}} d \eta \, \eta^{- \frac {1}{\epsilon}-1} \tilde
g_0 (\eta) \ e^{- \eta \delta \omega} \, .
\end{eqnarray*}
This clearly is of the same form as the anomalous contribution in our previous result (\ref{Malachit}). The
(unknown) function $\hat \beta_{0} (\zeta)$ is given by the integral involving the (unknown) function
$\tilde g_{0} (\eta)$. By construction the result (\ref{Tigereisen}) is valid for large $k$ and large $\tau$
and therefore picks up the structure of the singularities for $\tau \gg 1$. We conclude that the anomalous
contribution $e^{- \tau/\epsilon} \hat \beta_{0} (\zeta)$ is due to the singularities which approach
$\omega= - 1$, as claimed above.

We finally note that the leading singularity $\sim  (1 - \omega)^{1/\epsilon + \lambda}$ of the
eigenfunction $\beta_{\lambda} (\omega)$ implies that the Taylor coefficients of $e^{\lambda \tau}
\beta_{\lambda} (\omega)$ for large $k$ behave as
\begin{displaymath}
b_k (\tau) \sim (-1)^k \ k^{-\frac {1}{\epsilon} - 1} \left(\frac {e^\tau}{k} \right)^{\lambda} C_1 \left(1
+ {\cal O} \left( \frac{1}{k} \right) \right)\,,
\end{displaymath}
where $C_{1}$ is some constant. We thus recover the form (\ref{Tigereisen}) with $\tilde g_{0} (\eta) =
\eta^{-\lambda}$.

\subsection{Rigorous analysis of the limit $\tau \to \infty$} \label{rigorous analysis}

In the previous subsection we have argued that $\beta (\omega, \tau)$ for $\tau\to\infty$ tends to a function $\beta_\infty(\omega)$ that is analytic in any compact subset ${\cal K}$ of $\overline{\mathcal{U}}_\omega \setminus \{-1\}$. Furthermore, the eigenvalue analysis~\cite{PartI} as well as the results of subsections \ref{stronglyrigor}-\ref{asymptotic} suggest that within the linearized theory a perturbation for $\tau \to
\infty$ only leads to a constant shift of the circle. Assuming the existence of $\beta_{\infty} (\omega)$,
this can be proven rigorously.

We start from Eq.~(\ref{Beryll}): ${\mathcal L}_{\epsilon} \beta = 0$, rewritten as
\begin{eqnarray} \label{Menilit}
\left[ \left( 1 -  T^2 \right) \partial_T - \left( 1 - \omega^2 \right) \partial_\omega \right] \left( 1 +
\epsilon + \epsilon \, \omega \, \partial_\omega \right) \beta \left( \omega, \tau (T) \right)
\nonumber \\
= \left( 1 - \epsilon \right) \ \left( 1 + \omega^2 \right) \partial_\omega \, \beta\left( \omega, \tau (T)
\right) \,,
\end{eqnarray}
where $T= \tanh \tau/2$, (Eq.~\ref{Nigrin}), and we introduce the function
\begin{equation} \label{Marmor}
G (\omega, T) = \left( 1 + \epsilon + \epsilon \,  \omega \, \partial_\omega \right) \beta \left( \omega,
\tau (T) \right) \,.
\end{equation}
In terms of $G (\omega, T)$ the solution $\beta (\omega, \tau)$ regular at $\omega = 0$ is given by
\begin{equation} \label{Magnesit2}
\beta (\omega, \tau) = \frac {1}{\epsilon} \, \omega\,^{- 1/\epsilon - 1}
   \int\limits_0^\omega \omega\,'^{1/\epsilon} \, G (\omega\,', T (\tau)) d \omega\,'\,
\end{equation}
which generalizes Eq.~(\ref{Calcit}) to $\epsilon \not= 1$. We now write Eq.~(\ref{Menilit}) as
\begin{equation} \label{Jet}
\left[ \left( 1 - T^2 \right) \partial_{T} - \left( 1 - \omega^2 \right) \partial_{\omega} \right] G
(\omega, T)= H \left( \omega, T \right) \,,
\end{equation}
where
\begin{equation} \label{Iserin}
H (\omega, T) = \frac{1 - \epsilon}{\epsilon} \ \frac{1 + \omega^2}{\omega} \left[ G (\omega, T) - \frac{1 +
\epsilon}{\epsilon} \int\limits_{0}^1 x^{1/\epsilon} \ G (x \, \omega, T) d x \right]\,.
\end{equation}
Noting that $G (\omega, T) \equiv \zeta = (\omega + T)/(1 + \omega T)$ solves Eq.~(\ref{Jet}) for $H \equiv
0$ it is easily found that (\ref{Jet}) is equivalent to the integral equation
\begin{equation} \label {Inanga}
G (\omega, T) = G (0, \zeta) - \int\limits_{0}^\omega \frac{1}{1 - \omega\,'^2} \ H \left( \omega\,', \
\frac{\zeta - \omega\,'}{1 - \omega\,' \zeta} \right) d \omega\,' \,.
\end{equation}
We now define
\begin{equation} \label{Hayn}
\Delta (\omega, T) = G (\omega, T) - G (0, T) \,.
\end{equation}
Eq.~(\ref{Inanga}) yields
\begin{eqnarray} \label{Idokras}
\Delta (\omega, T) = G (0, \zeta) - G (0, T) - \, \frac{1 - \epsilon}{\epsilon} \int\limits_{0}^\omega
\frac{1}{\omega\,'} \ \frac{1 + \omega\,'^2}{1 - \omega\,'^2} \Bigg[ \Delta \left( \omega\,', \frac{\zeta -
\omega\,'}{1 - \omega\,'  \zeta} \right) \nonumber \\ - \, \frac{1 + \epsilon}{\epsilon} \int\limits_0^1
x^{1/\epsilon} \Delta \left( x \, \omega\,', \ \frac{\zeta - \omega\,'}{1 - \omega\,' \zeta}  \right) d x
\Bigg] d \omega\,' \,,
\end{eqnarray}
where we have written out $H$ explicitly. In view of the results of subsect\,\ref{smoothness} we now assume
that $\lim\limits_{T \to 1} G (\omega,  T)$ exists for $\omega \, \in\, {\cal K}$. We further
note that for $T \to 1$ and $\omega \not= - 1$, both $\zeta$ and $(\zeta - \omega\,') /(1 - \omega\,'
\zeta)$ tend to $1$. Eq.~(\ref{Idokras}) reduces to the homogenous integral equation
\begin{equation} \label{Lasurit}
\Delta (\omega, 1) = - \frac{1 - \epsilon}{\epsilon} \int\limits_0^\omega \frac{1}{\omega\,'} \ \frac{1 +
\omega\,'^2}{1 - \omega\,'^2} \Bigg[ \Delta (\omega\,',1) - \frac{1 + \epsilon}{\epsilon} \int\limits_0^1
x^{1/\epsilon}\  \Delta \left( x \, \omega\,', 1 \right) d x \Bigg] d \omega\,' \,.
\end{equation}
It is easily checked that for all $\epsilon > 0$ the only solution of (\ref{Lasurit}) analytic in a
neighborhood of $\omega = 0$ is the trivial one:
\begin{equation} \label{Keratit}
\Delta (\omega, 1) \equiv 0 \,.
\end{equation}
To see this, we assume that the Taylor expansion of $\Delta (\omega, 1)$ starts with a lowest order term
$a_{k} \, \omega^k$, $k\ge 1, \ a_k \not= 0$. Eq.~(\ref{Lasurit}) yields $a_{k} = 0$, contradicting our
assumption.

We thus have shown that provided $G(\omega, 1)$ exists and is analytic for $\omega \, \in \,
{\cal K}$, the only solution to our problem is
\begin{equation} \label{Geode}
G (\omega, 1) \equiv G (0, 1) \,,
\end{equation}
implying
\begin{equation} \label{Glaslava}
\beta_{\infty} (\omega) = \frac{G (0, 1)}{1 + \epsilon} \,,
\end{equation}
which for $\epsilon = 1$ reduces to Eq.~(\ref{Florstein}).

\subsection{Numerical illustration} \label{numerical}

In this section, we show numerical results of the linear evolution. We approximately solve the PDE
(\ref{Biotit}) by truncating the series expansion (\ref{Serpentin}),
\[
\beta=\sum_{k=0}^{\infty}b_{k}\omega^{k},\] at $k=N$. The ODE system for $b_{k} (\tau)$ has been given
in Eq.~(\ref {Sugilith})
\begin{eqnarray*}
2 \partial_{\tau} b_{0} &=& {\frac{2+\epsilon} {1+\epsilon}} \ b_{1} \, ,
\\
2 \partial_{\tau} b_{k} &=& \frac {k+1}{1 + \epsilon + \epsilon k} \Bigl [ (2 + \epsilon + \epsilon k) \
b_{k + 1} - \epsilon (k- 1) b_{k-1} \Bigr ] \quad\mbox{for } k \ge 1 \,. \nonumber
\end{eqnarray*}
With the $b_k(0)$ given by the initial condition, the $b_k(\tau)$ can be determined recursively by the
Runge-Kutta time stepping method. We choose the cut-off $N = 2000$ in the simulation. Adaptive time steps
are chosen which ensure that the difference between 4-th order and 5-th order Runge Kutta methods is within
$10^{-15}$. In the sequel we present results for
\begin{eqnarray} \label{Tuerkis}
\delta \beta (\omega, \tau) & =  & \beta (\omega, \tau) - \beta (0, \tau).
\end{eqnarray}
The subtraction eliminates the overall shift of the evolving body.

\begin{figure}[h]
\begin{center}
\subfigure[$\epsilon=1/10$]{\includegraphics[scale=0.85]{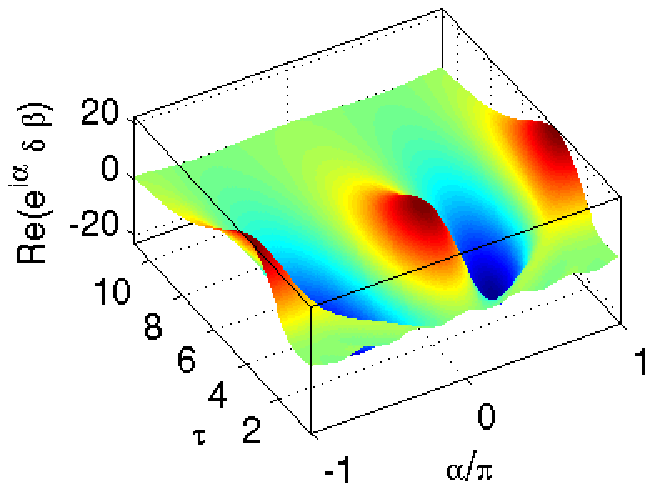}}
\subfigure[$\epsilon=1/2$]{\includegraphics [scale=0.85]{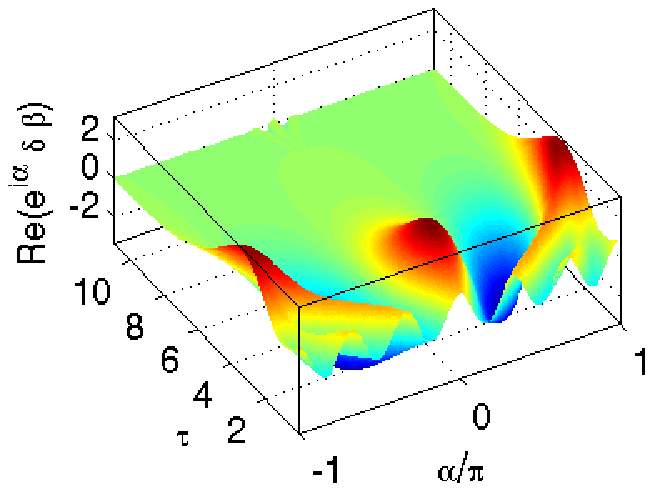}}
\\
\subfigure[Times 0, 0.5, 2, 6 for $\epsilon=1/10$ ]{\includegraphics [scale=0.60] {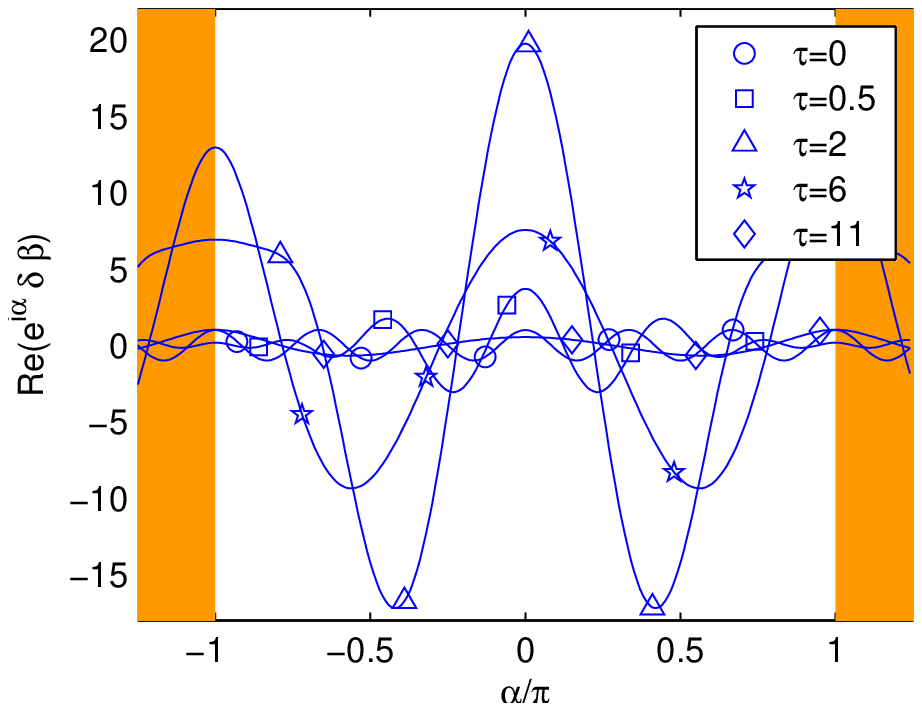}}
\subfigure[Same time steps for $\epsilon=1/2$ ]{\includegraphics [scale=0.60] {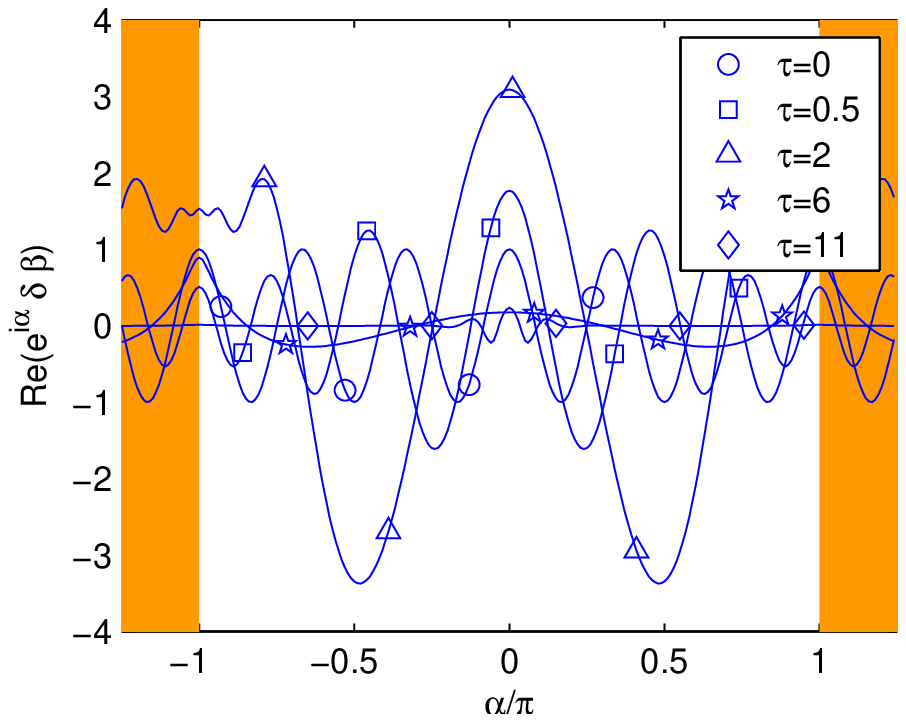}}
\caption{Evolution of
${\rm Re}\big[\omega \,\delta \beta (\omega, \tau)\big]$ for $\omega=e^{i\alpha}$ for the initial condition
(\ref{Turmalin}). a) Overview plot for $\epsilon=1/10$ and times $0\le \tau\le 11$. b) The same for
$\epsilon = 1/2$ and times $0\le \tau\le 11$. c) Detailed data for time steps 0, 0.5, 2, 6, 11 for
$\epsilon=1/10$; the angle is normalized as $\alpha/\pi$; orange areas are overlap regions plotted to make
the structure at the back visible. d) The same for $\epsilon=1/2$. } \label{fig4}
\end{center}
\end{figure}

We first present results typical for a delocalized initial condition, choosing
\begin{equation} \label{Turmalin}
\beta (\omega, 0) = \omega^5 \,.
\end{equation}
Fig.~\ref{fig4} shows the evolution of ${\rm Re} \Big[ \omega \,\delta \beta (\omega, \tau) \Big]$,
$\omega = e^{i \alpha}$, with $\epsilon = 1/10$ or $\epsilon = 1/2$,
respectively. In physical space, ${\rm Re} [\omega \,\delta \beta]$ is the component of the perturbation
normal to the unperturbed but shifted circle at angle $- \alpha$. Panels a and b show that the qualitative
behavior is quite similar for both values of $\epsilon$ shown. Panels c and d give a detailed view on the
state for several time steps; here an extended range of $\alpha$ is shown, so that the behavior both at
$\alpha = 0$ and $|\alpha| = \pi$ is clearly seen. For small times the perturbations increase in the front
half $|\alpha| < \pi/2$ of the circle and decrease in the back half. The maximum at $\alpha = 0$ increases
and broadens strongly, whereas the other perturbations are shifted towards $\alpha = \pm \pi$. At later
times the perturbations decrease at the front half, while at the back a transient increase is observed which
is due to the advection of the dynamically generated large amplitude of the perturbation towards $\alpha = \pi$. The results for
$\epsilon=1/10$ or $\epsilon=1/2$ essentially differ only in two respects. First, for $\epsilon = 1/10$ the
perturbation at intermediate times is amplified much more than for $\epsilon = 1/2$. Second, for $\tau = 2$
and $\epsilon = 1/2$, remainders of individual maxima that initially are located at $\alpha\ne0$, still can be seen near $\alpha = \pm \pi$ (this structure was very pronounced for $\epsilon=1$ as
discussed in \cite{eber07}), whereas for $\epsilon = 1/10$ this structure is completely damped out and
yields a broad maximum.

\begin{figure}[t]
\begin{center}
\subfigure[$\epsilon=1/10$]{\includegraphics[scale=0.85]{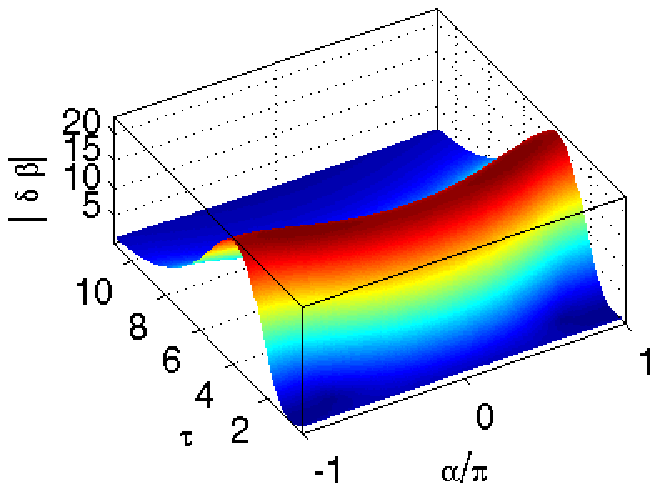}}
\subfigure[$\epsilon=1/2$]{\includegraphics [scale=0.85]{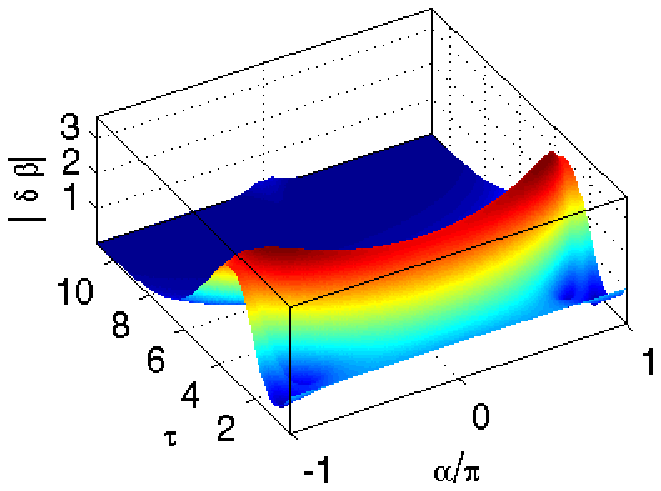}} \caption{Evolution of
$|\delta \beta (\omega, \tau)|$ for the initial condition (\ref{Turmalin}) with (a) $\epsilon = 1/10$ and  (b) $\epsilon=1/2$. } \label{fig5}
\end{center}
\end{figure}

Fig.~\ref{fig5} shows $|\delta \beta|$ as a function of $\alpha$ and $\tau$. It illustrates how the maximum
of the absolute value of the perturbation is advected towards $\alpha = \pi$, where it decays. For $\epsilon
= 1/10$ the behavior of $|\delta \beta|$ is quite smooth, whereas for $\epsilon = 1/2$ some small scale
structure is observed near $\alpha = \pm\pi$.

\begin{figure}[h]
\begin{center}
\subfigure[$\epsilon=1/10$]{\includegraphics[scale=0.6]{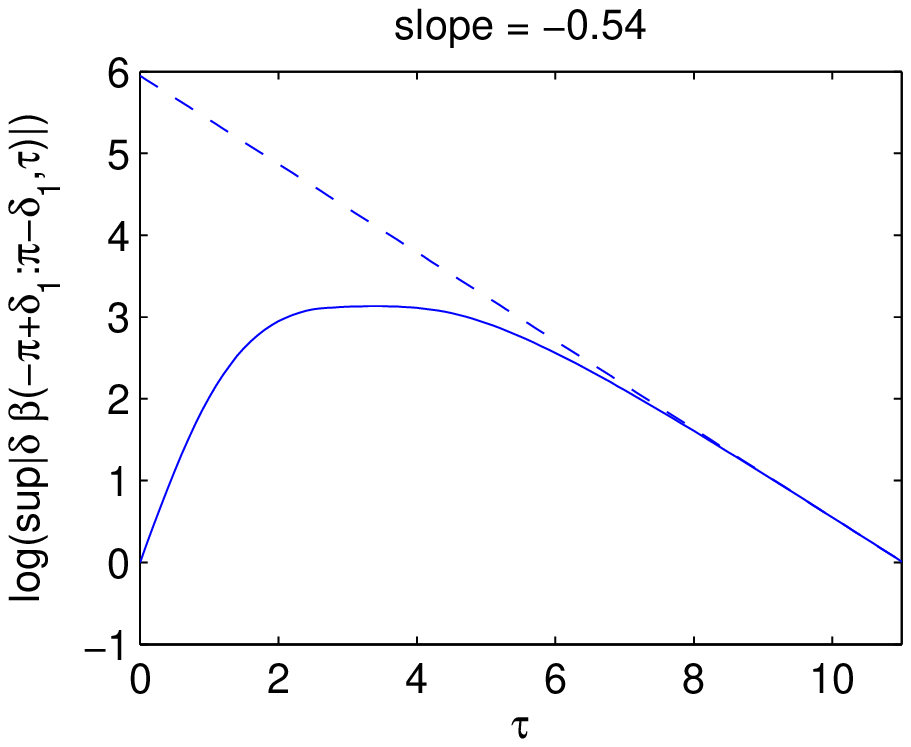}}
\subfigure[$\epsilon=1/2$]{\includegraphics [scale=0.6] {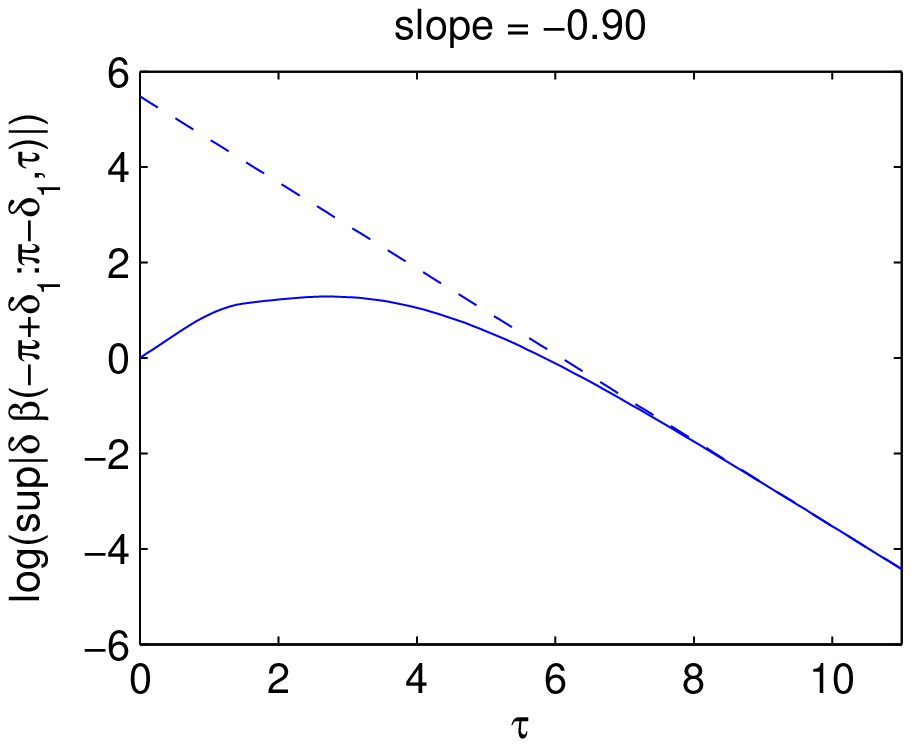}} \caption{$\log
\sup_{\alpha\in[-\pi+\delta_1,\pi-\delta_1]}|\delta\beta(\alpha,\tau)|$, $\delta_1 = \pi/8$ as a function of
$\tau$ for the data presented in Fig.~\ref{fig5}. a) $\epsilon=1/10$, the line has slope $\lambda_{1} (1/10)
= - 0.54$. b) $\epsilon=1/2$, the line has slope $\lambda_{1} (1/2) = - 0.90$.} \label{fig6}
\end{center}
\end{figure}

Outside a neighborhood of $\alpha = \pi$ we expect to see asymptotically exponential relaxation: $\delta
\beta \sim e^{\lambda_1  \tau}$. From the results given in paper I~\cite{PartI} in Fig.~\ref{fig1}, we
expect $\lambda_{1} \approx - 0.546$ for $\epsilon = 1/10$, and $\lambda_{1} \approx - 0.905$ for $\epsilon
= 1/2$, respectively. These predictions are tested in Fig.~\ref{fig6}. Since the maximum of $|\delta \beta|$
advects along the circle, we plot $\ln |\delta \beta_{\rm max} (\tau)|$ as a function of $\tau$, where
\[
|\delta \beta_{\rm max} (\tau)| = \sup_{\alpha \epsilon [\pi + \delta_1, \pi - \delta_1]}
   |\delta \beta (e^{i \alpha}, \tau)| \,.
\]
We choose $\delta_1 = \pi/8$. For smaller values of $\delta_{1}$ it needs larger values of $\tau$ to reach
the asymptotic behavior. We fit the curve for data at $10\le \tau \le 11$. As Fig.~\ref{fig6} illustrates, the expected asymptotic behavior is observed.

\begin{figure}[t]
\begin{center}
\subfigure[$\epsilon=1/10$]{\includegraphics[scale=0.85]{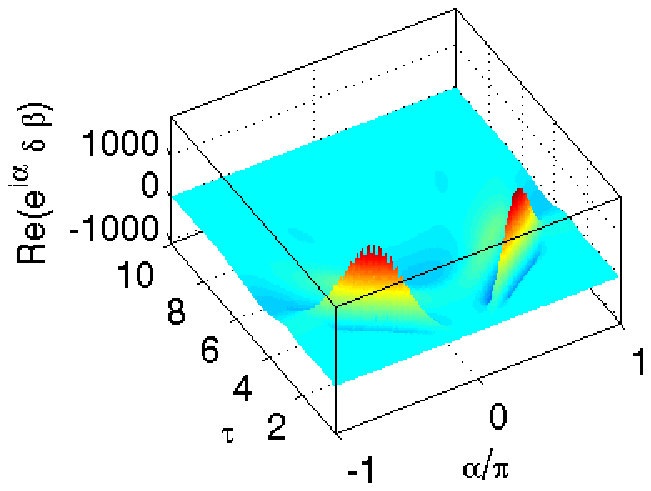}}
\subfigure[$\epsilon=1/2$]{\includegraphics[scale=0.85]{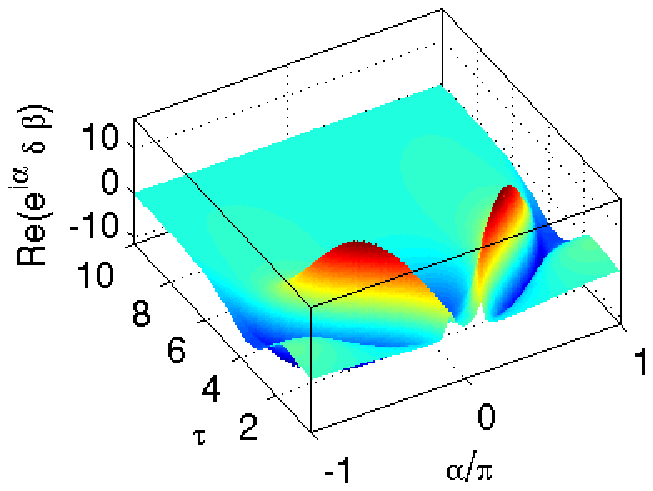}}\\
\subfigure[$\epsilon=1/10$]{\includegraphics[scale=0.60]{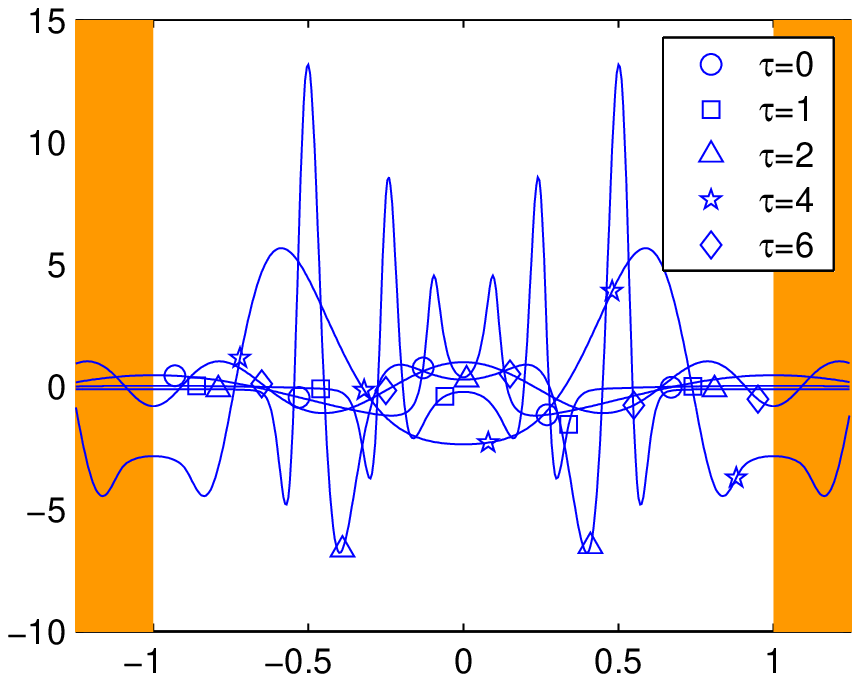}}
\subfigure[$\epsilon=1/2$]{\includegraphics[scale=0.60]{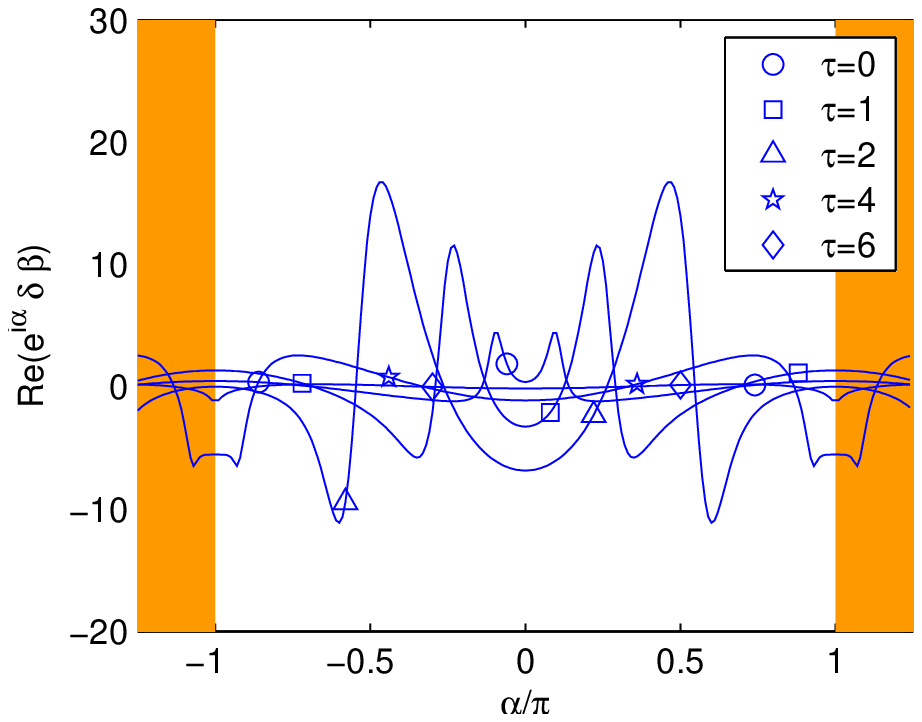}}\\
\caption{Evolution of ${\rm Re}\big[\omega\,\delta\beta(\omega,\tau)\big]$ for the initial condition~(\ref{Thumalin}) with $\gamma = 1.1 \, e^{- i \pi/10}$ for $\epsilon=1/10$ and $\epsilon = 1/2$. In panel c), Re($ \omega \delta \beta$) is scaled by factors $\alpha_0 (\tau)$, where $\alpha_0 (0) = 1$, $\alpha_0 (1) =
0.03$, $\alpha_0 (2) = 0.007$, $\alpha_0 (4) =0.025$, $\alpha_0 (6)=0.05$.} \label{fig7}
\end{center}
\end{figure}

\begin{figure}[t]
\begin{center}
\subfigure[$\epsilon=1/10$]{\includegraphics [scale=0.85] {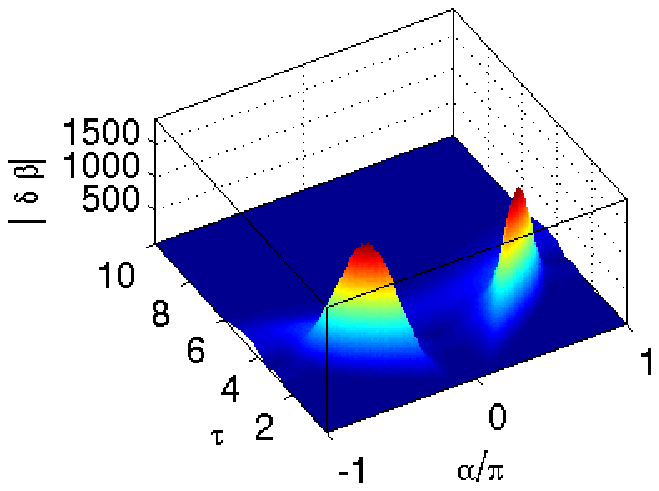}}
\subfigure[$\epsilon=1/2$]{\includegraphics [scale=0.85] {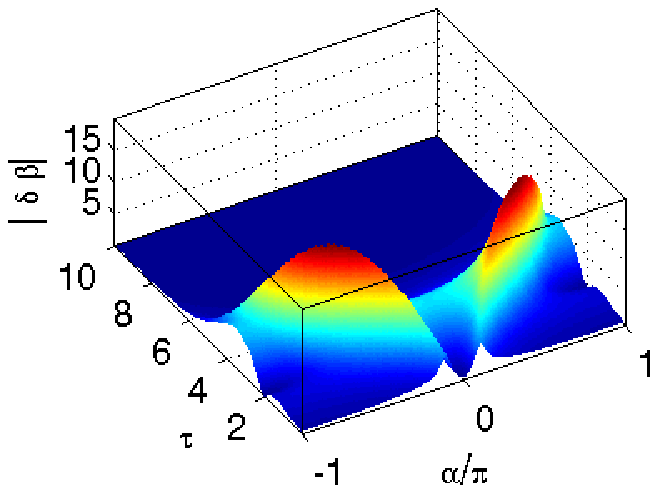}}
\caption{The same evolution as in Fig.~\ref{fig7}, plotted is now $|\delta \beta (\alpha, \tau)|$ as a function of $\alpha$ and $\tau$.
Left column: $\epsilon=1/10$, right column: $\epsilon=1/2$. Note the large difference in the scale of $|\delta\beta|$, reflecting the strong $\epsilon$ dependence of the amplitude.} \label{fig8}
\end{center}
\end{figure}

We now consider a more localized initial condition:
\begin{eqnarray}\label {Thumalin}
\beta (\omega, 0) = \frac{1}{2} \left( \frac{\gamma}{\gamma - \omega}
    + \frac{\gamma^*}{\gamma^* - \omega} \right)
    = \frac12 \sum_{j=0}^\infty\left[\left(\frac\omega\gamma\right)^j+\left(\frac\omega{\gamma^*}\right)^j\right].
\end{eqnarray}
With the choice $\gamma = 1.1 e^{i \pi/10}$ it shows two fairly sharp peaks centered symmetrically close to $\alpha = 0$. Similarly to Fig.~\ref{fig4}, Fig.~\ref{fig7} shows ${\rm Re} (\omega \delta \beta (\omega, \tau))$ for $\epsilon=1/10$ and $\epsilon = 1/2$. For $\epsilon=1/10$ we rescaled the amplitude in panel c by a time dependent factor $a_0 (\tau)$ in order to show all curves in the same plot. Panels a and b show that the time dependent shift of the structure is essentially independent of $\epsilon$, implying that advection is determined by the automorphism $\zeta=\zeta (\omega, T(\tau))$. Panels c and d illustrate that also the detailed structure at given time $\tau$ is fairly independent of $\epsilon$, but for $\epsilon=1/10$ the amplitude at intermediate times is enhanced much more than for $\epsilon=1/2$ (cf.\ the rescaling factors $a_0(\tau)$ given in the figure caption). Fig.~\ref{fig8} shows $|\delta \beta|$ as function of $\alpha$ and $\tau$, similarly to Fig.~\ref{fig5}. Again the advection of the maximum towards $\alpha=\pi$, its increase as long as it is in the front half, and its final decay in the back half are clearly seen.

In summary, all numerical results presented here and in previous subsections support our analysis.


\section{Nonlinear Evolution} \label{Sec:Nonlin}

In Sect.~\ref{analysis} we discussed the solution of the linearized evolution equation. Here we seek to
determine the effect of the nonlinearity. In Subsect.~\ref{circular shape} we consider small perturbations
of the circle. Subsect.~\ref{GenShapes} presents examples of the evolution of more general initial shapes.

To calculate the nonlinear evolution in a large range of time is difficult. Using a Fourier representation
of the interface it for large $\tau$ needs wave numbers of order $e^{\tau} \gg 1$ to resolve the collapsing
region near $\omega = - 1$. Modes of large wave number can also be expected to play an important role at the
front part of the bubble. Approximating a small region near $\omega = + 1$ as planar, we may invoke well
known results~\cite{man04,gianne08,How92} on the instability of a planar interface: in linear approximation the amplitude of a Fourier mode of wave number $k$ increases like $e^{s (k) \tau}$, where $$s(k)=\frac k {1+\epsilon k}.$$ Thus with the present regularization all Fourier modes are unstable, whereas with curvature regularization only a finite unstable band exists. The strong increase of a perturbation localized near $\omega=+1$, as discussed in Subsect.~\ref{strongly}, can be considered to result from this instability of modes $k \gg 1/\epsilon$. Nonetheless, despite stringent demands on resolution and time steps, we believe that the results presented here exemplify the nonlinear effects. The numerical methods used to solve the nonlinear equations (\ref{Girasol}) and (\ref{Azurit}) are summarized in the Appendix. We use a Fourier representation with cutoff $k_{max}=N$, and we solve the resulting system of ordinary differential equations with a 4th order Runge-Kutta method with time step $\Delta t$. ($N$ and $\Delta t$ are given in the figure captions.) The numerics abruptly breaks down at some time $t_{\rm max}(\epsilon)$. The time range shown in the figures therefore depends both on $\epsilon$ and on the initial condition.


\subsection{Small perturbations of the circle} \label{circular shape}

We here consider perturbations $\eta\,\beta (\omega, 0)$ of the circle, with $\eta \ll 1$. In order to
compare with the linear evolution we define the nonlinear counterpart to $\delta \beta (\omega, \tau) =
\beta (\omega, \tau) - \beta (0,\tau)$ as
\begin{equation} \label{nonleq_10}
\delta \beta_{nl} (\omega, \tau) = \frac{\hat f (\omega, t) - \hat f (0, t)}\eta ,
\end{equation}
where $t = \frac{1+\epsilon}{2} \tau$, Eq.~(\ref{Blauquarz}), and $\hat f (\omega, t)$ is defined in
Eq.~(\ref{Apatit}). For $\eta \to 0$, $\delta \beta_{nl}$ reduces to $\delta \beta$.

\begin{figure}
\begin{center}
\subfigure[$\epsilon=1/10$, $\omega=1$]{\includegraphics[scale=0.6]{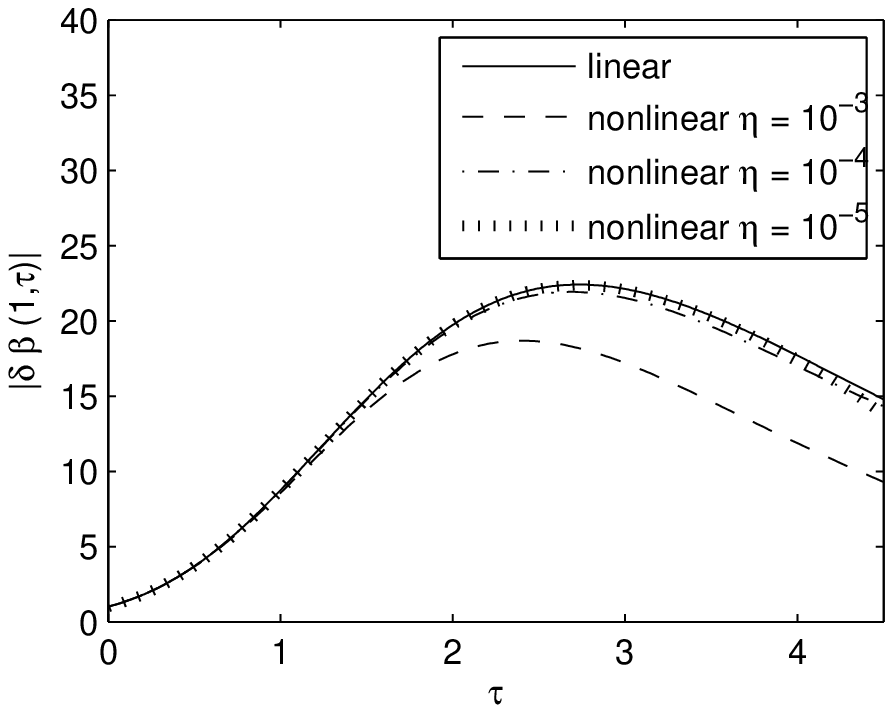}}
\subfigure[$\epsilon=1/10$, $\omega=-1$]{\includegraphics[scale=0.6]{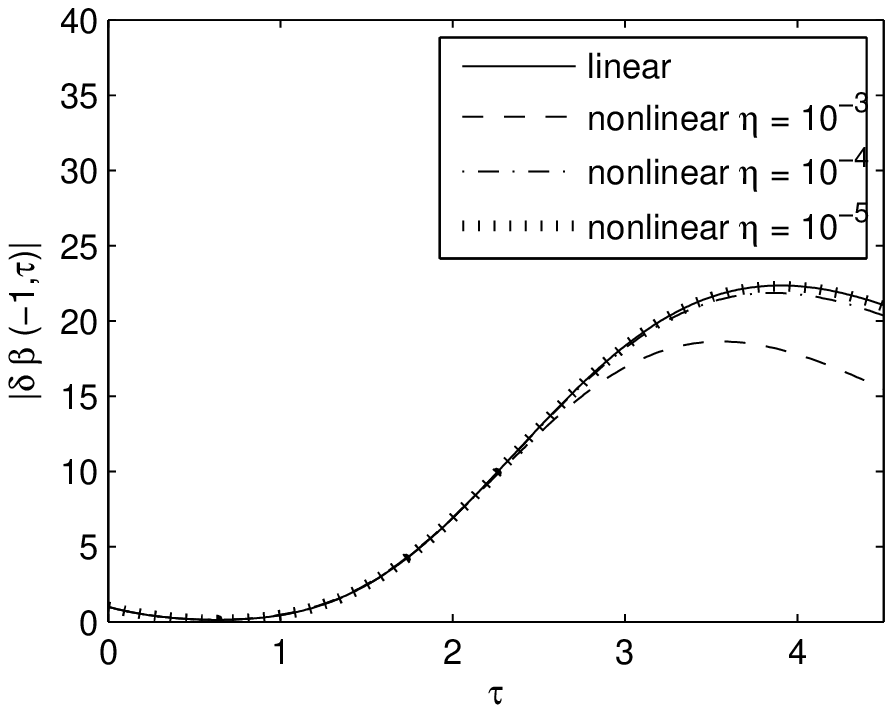}}\\
\subfigure[$\epsilon=1/2$, $\omega=1$]{\includegraphics[scale=0.6]{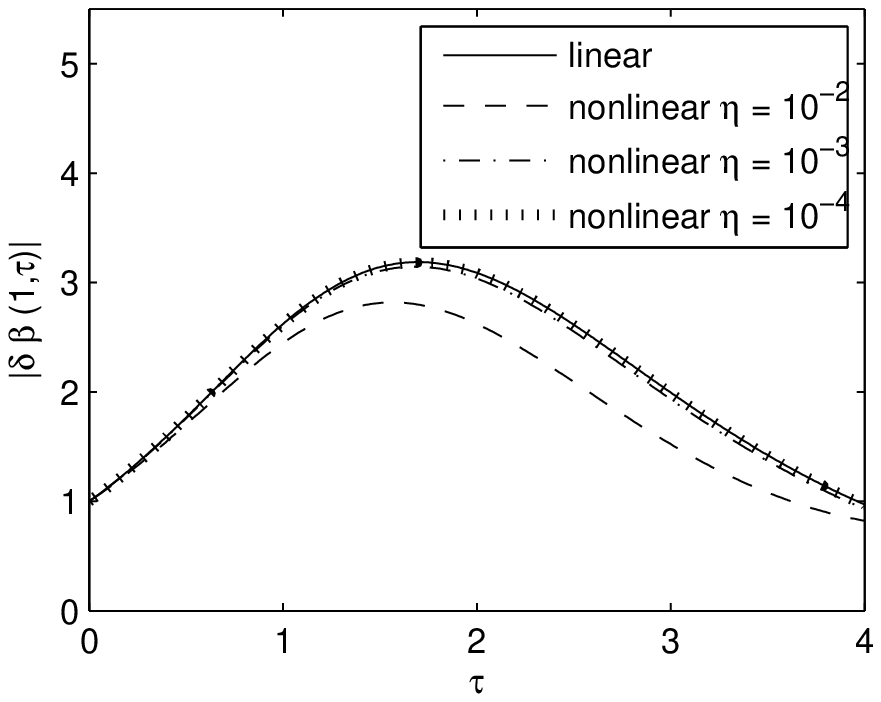}}
\subfigure[$\epsilon=1/2$, $\omega=-1$]{\includegraphics[scale=0.6]{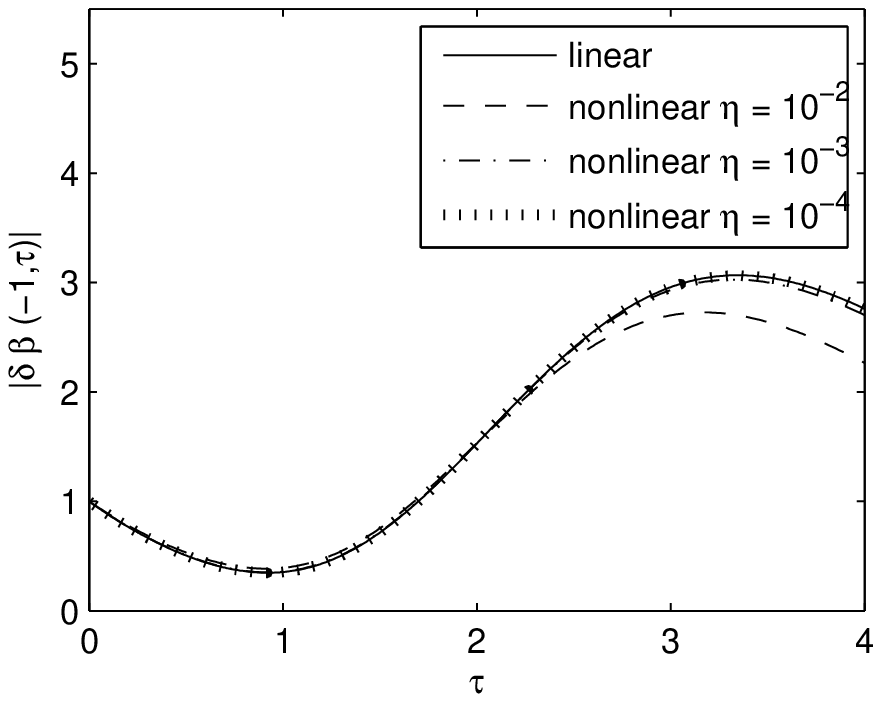}}
\caption{Initial condition: $\beta_{nl}(\omega,0)=\omega^5$, plotted are $\delta \beta (1, \tau)$ and $\delta \beta (-1, \tau)$ as a function of $\tau$ for $\epsilon=1/10$ and $\epsilon=1/2$ for different $\eta$.
 ($N=256$, $\Delta t=0.001$)}\label{fig10N}
\end{center}
\end{figure}

We first consider the delocalized initial condition (\ref{Turmalin}): ${\hat f} (\omega, 0) = \eta\,
\omega^5$. Even for
very small $\eta$ it is not obvious a priori that the nonlinearity is unimportant. As recalled above,
perturbations at the front may increase dramatically, and the collapsing region at the back, where an
eigenmode expansion is bound to fail, also might be quite sensitive to nonlinear effects.
We therefore in Fig.~\ref{fig10N} show $ \delta\beta_{nl}(+1,\tau)$ and $ \delta
\beta_{nl}(-1,\tau)$ for $\epsilon=\frac{1}{10}$ or $\frac{1}{2}$ and
several values of $\eta$. It is seen that for very small values of $\eta$
the nonlinear theory essentially reproduces the results of the linear
approximation. Deviations outside some initial time range become visible
for $\eta\ge10^{-4}$, $(\epsilon=\frac{1}{10})$, or $\eta\ge10^{-3}$, $(\epsilon=
\frac{1}{2})$, respectively, but even then the shape of the curves is
similar to the linear approximation. This suggests that also in the
forward and backward regions the nonlinearity for small perturbations
does not qualitatively change the results of the linear approximation.

\begin{figure}
\begin{center}
\subfigure[$\epsilon=1/10$, $\eta = 10^{-3}$]{\includegraphics[scale=0.6]{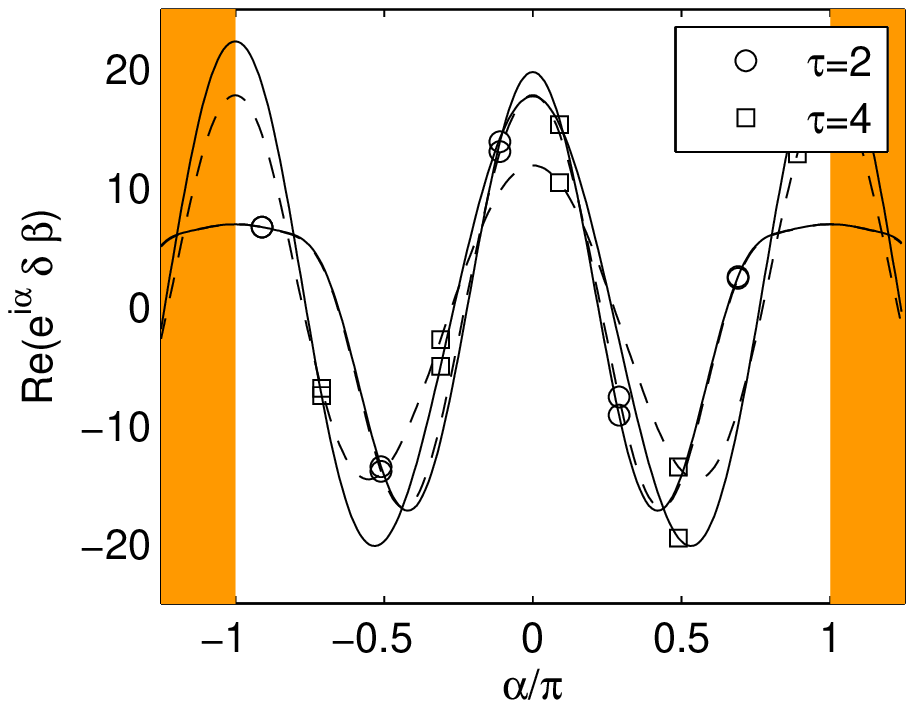}}
\subfigure[$\epsilon=1/2$, $\eta = 10^{-2}$]{\includegraphics[scale=0.6]{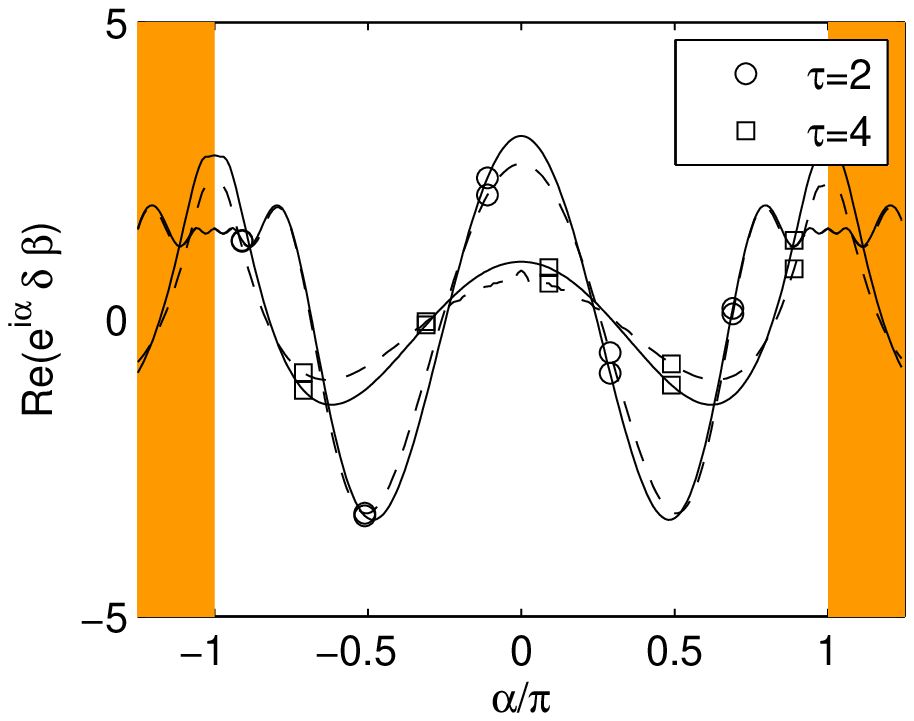}}
\caption{Evolution of
${\rm Re}\big[\omega \,\delta \beta (\omega, \tau)\big]$ for $\omega=e^{i\alpha}$ for the initial condition
(\ref{Turmalin}) for $\epsilon=1/10$ and $\epsilon=1/2$ at different $\tau$. ($N=256$, $\Delta t=0.001$)}\label{fig10N1}
\end{center}
\end{figure}

The results shown in Fig.~\ref{fig10N1} support this conclusion. We here plot
$Re(e^{i \alpha} \delta \beta_{nl}(e^{i \alpha},\tau)$ as function of
$\alpha/\pi$, for values of $\tau$ where a deviation from the linear
approximation is visible. We observe that the nonlinearity essentially
influences the amplitude but not the shift of the perturbation. The
overall structure is most similar to the linear approximation.

\begin{figure}
\subfigure[$\epsilon=1/10$]{\includegraphics[scale=0.6]{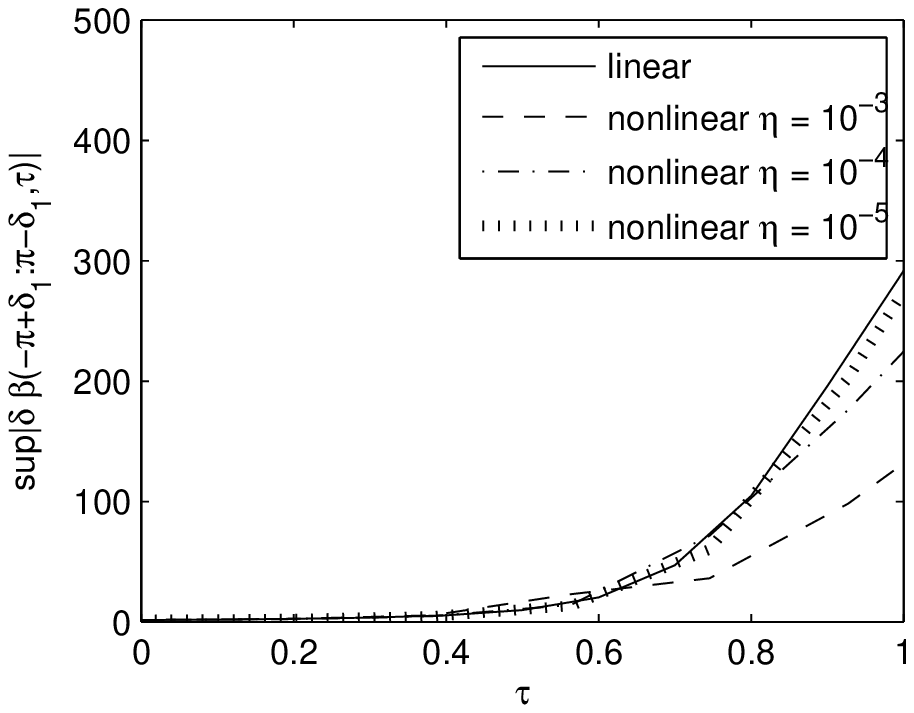}}
\subfigure[$\epsilon=1/2$]{\includegraphics[scale=0.6]{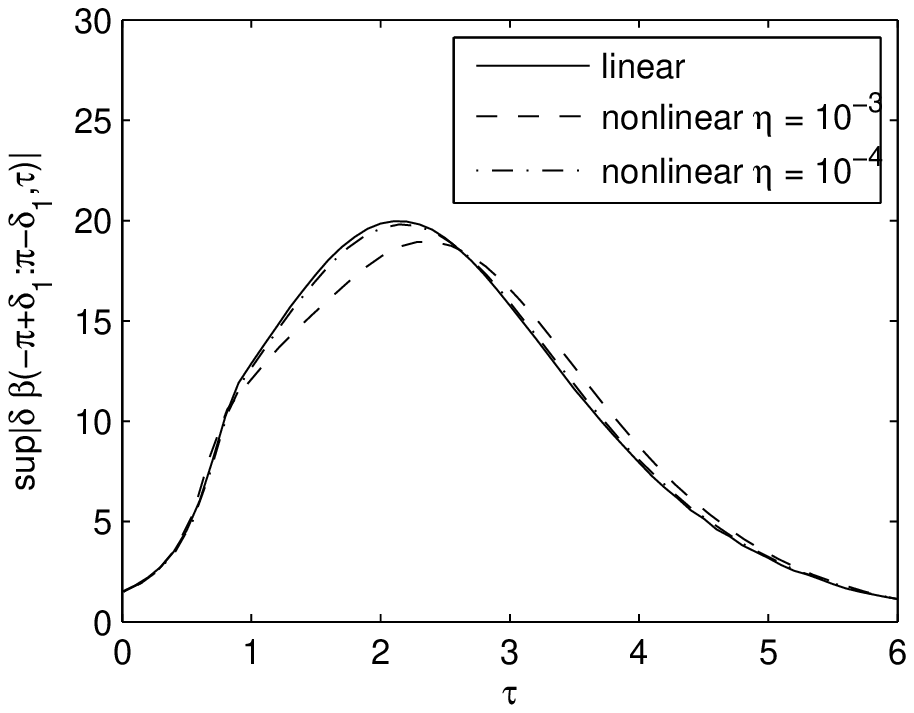}}\\
\subfigure[$\epsilon=1/10$, $\eta = 10^{-3}$] {\includegraphics[scale=0.6]{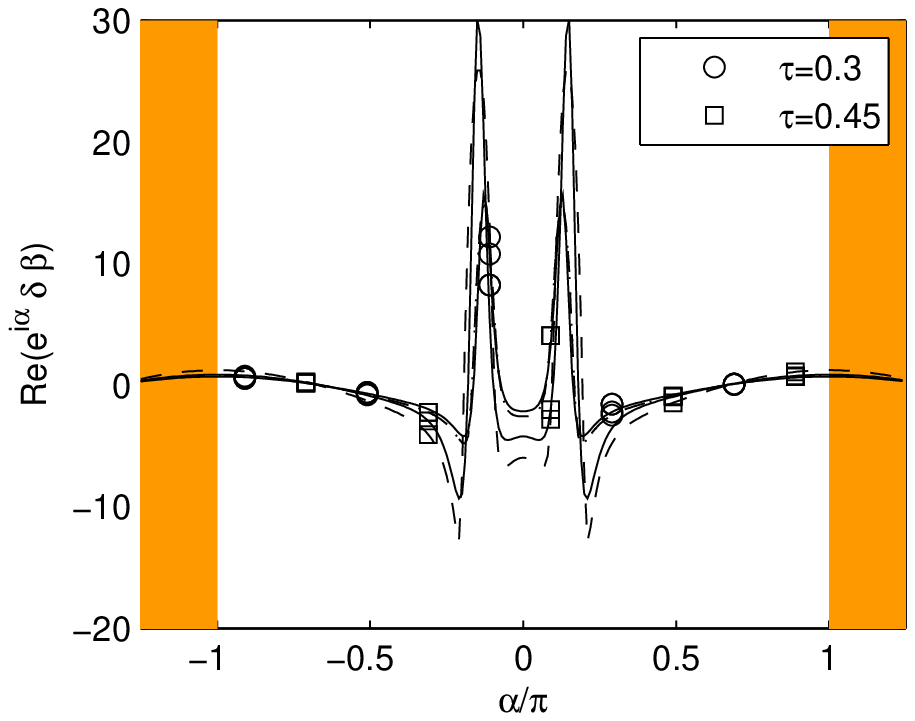}}
\subfigure[$\epsilon=1/2$, $\eta = 10^{-3}$]{\includegraphics[scale=0.6]{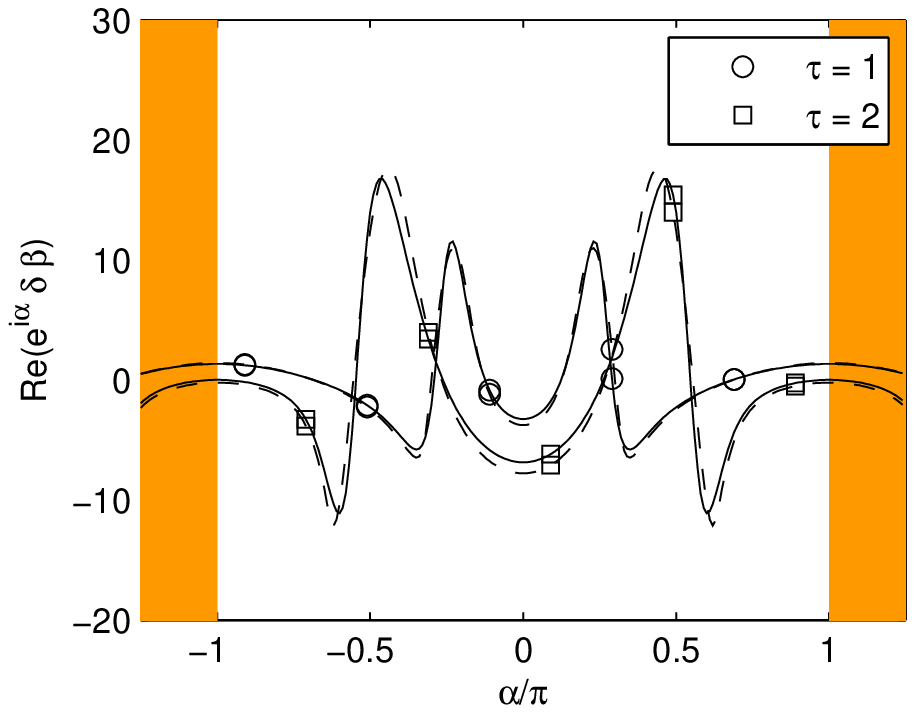}}
\caption{ Initial condition: $\beta(\omega,0)=\frac{1}{2}
\frac{\gamma}{\gamma-\omega}+\frac{1}{2} \frac{\gamma^\ast}{\gamma^
\ast-\omega}$ , plotted are $\sup_{\alpha\in[-\pi+\delta_1,\pi-\delta_1]}|\delta\beta(\alpha,\tau)|$, $\delta_1 = \pi/8$ as a function of $\tau$ for
(a)$\epsilon=1/10$ and (b)$\epsilon=1/2$ for different $\eta$. Evolution of
${\rm Re}\big[\omega \,\delta \beta (\omega, \tau)\big]$ for $\omega=e^{i\alpha}$ for (c)$\epsilon=1/10$ and (d)$\epsilon=1/2$ at different $\tau$. ($N=256$, $\Delta t=0.001$)} \label{fig12N}
\end{figure}

Such results are also found for other delocalized perturbations of type
$\eta \omega^n$. Also more localized perturbations behave similar. For
the initial condition (\ref{Thumalin}), $\beta(\omega,0)=\frac{1}{2}
\frac{\gamma}{\gamma-\omega}+\frac{1}{2} \frac{\gamma^\ast}{\gamma^
\ast-\omega}$ this is illustrated in Fig.~\ref{fig12N}. We again observe that the
nonlinearity essentially influences only the amplitude of the
perturbation, but leaves the qualitative structure almost unchanged.
All these results suggest that the circle is the asymptotic attractor
for weak perturbations.

\begin{figure}
\subfigure[$\epsilon=1/10$, nonlinear]{\includegraphics[scale=0.6]{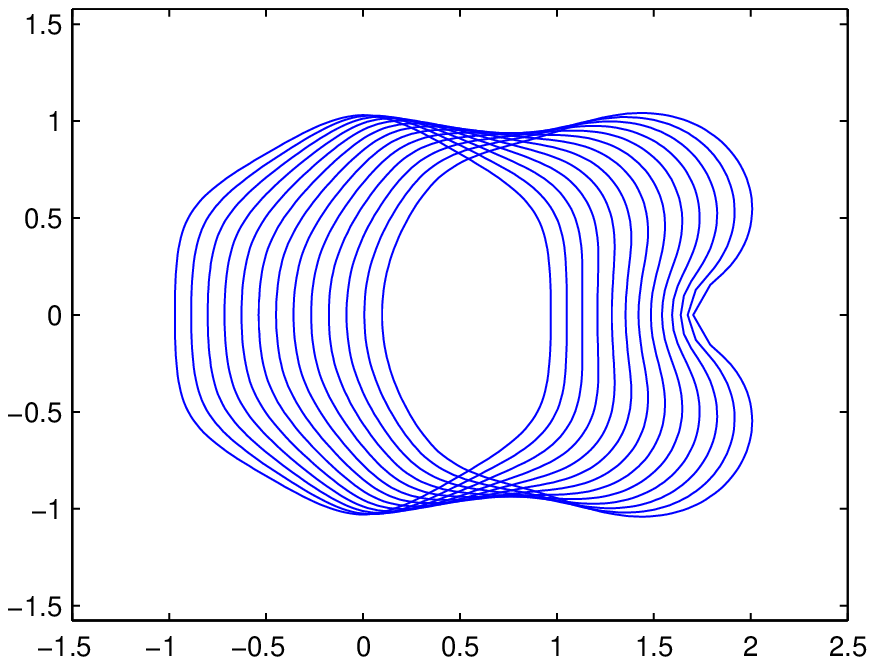}}
\subfigure[$\epsilon=1/10$, linear]{\includegraphics[scale=0.6]{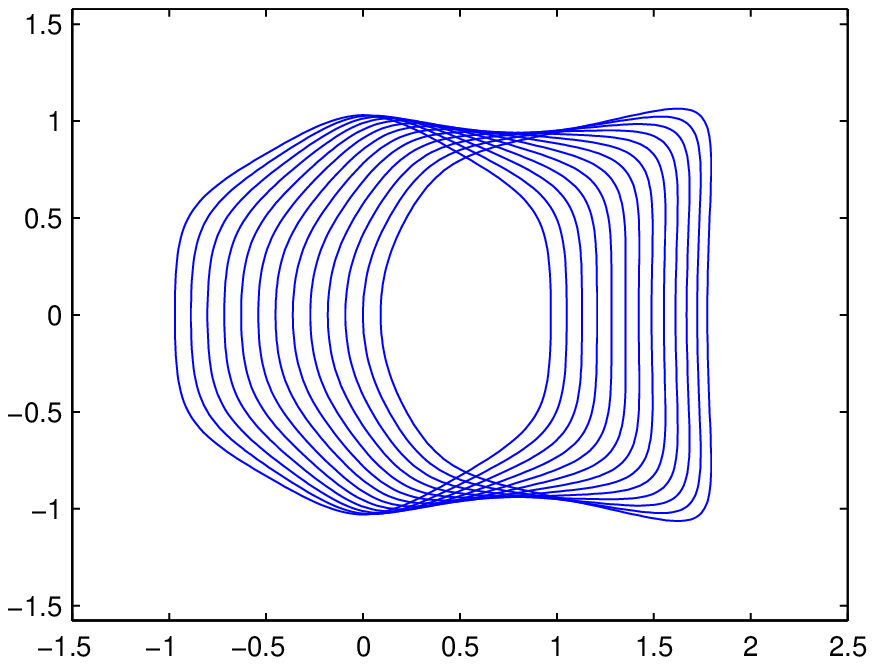}}
\subfigure[$\epsilon=0$, nonlinear]{\includegraphics[scale=0.6]{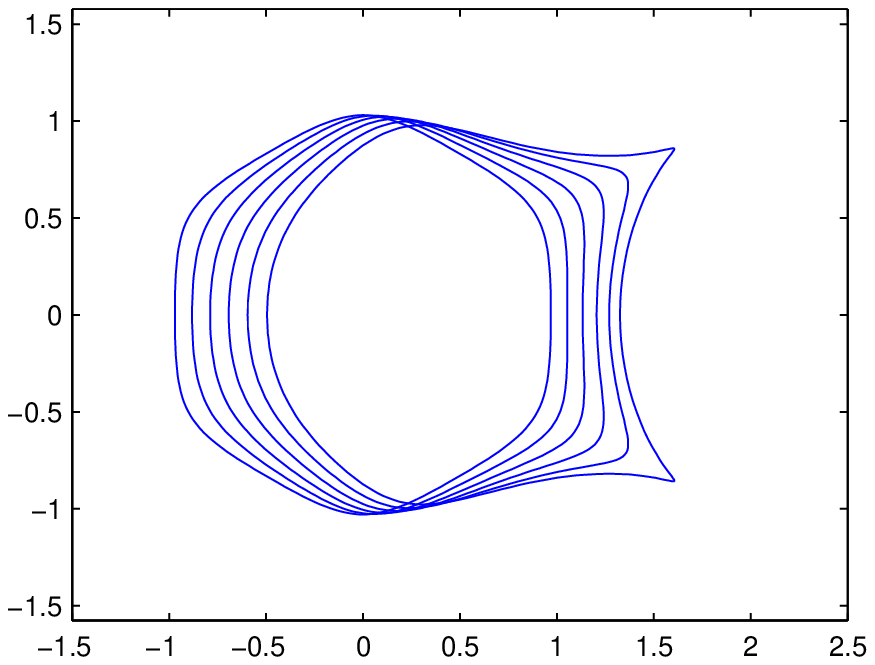}}
\caption{Initial condition: $\beta(\omega,0)=-0.03\;\omega^5$. (a) and (b) show the nonlinear and the linear evolution respectively for $\epsilon=1/10$.
(c) shows the nonlinear evolution for $\epsilon=0$. ($N=512$, $\Delta t=0.00025$)} \label{fig11N}
\end{figure}

For larger initial perturbations it is unlikely that the circle is
recovered asymptotically. Rather we may observe branching. This is
illustrated in Fig.~\ref{fig11N} with the initial condition $\beta(\omega,0)=-0.03\;
\omega^5$. Panel (a) shows snapshots of the interface in physical space
$z=x+i y=f(\omega,t)$, with $\epsilon=\frac{1}{10}$, as resulting from
the nonlinear evolution. For comparison panel (b) shows the linearized
evolution, and panel (c) shows the result of the unregularized model
$\epsilon=0$. Snapshots are taken at times $t= 0.05 n$, where $n=0,1,\ldots, 12$ in panels (a) and (b), and $n=0,1,\ldots, 5$ in panel (c). Clearly the cusps which
in the unregularized model occur for $t\approx 0.25$ for $\epsilon=
\frac{1}{10}$ are suppressed both according to the linear and the
nonlinear evolution. A qualitative effect of the nonlinearity is
observed for $t>0.1$. Whereas the linear approximation develops shoulders
connected by some flat part of the interface, the nonlinear evolution
results in two branches separated by a valley. Since the bottom of the
valley moves slower than the tips of the branches, the valley is likely to
evolve into a deep fjord.

\begin{figure}[t]
\subfigure[]{\includegraphics[scale=0.6]{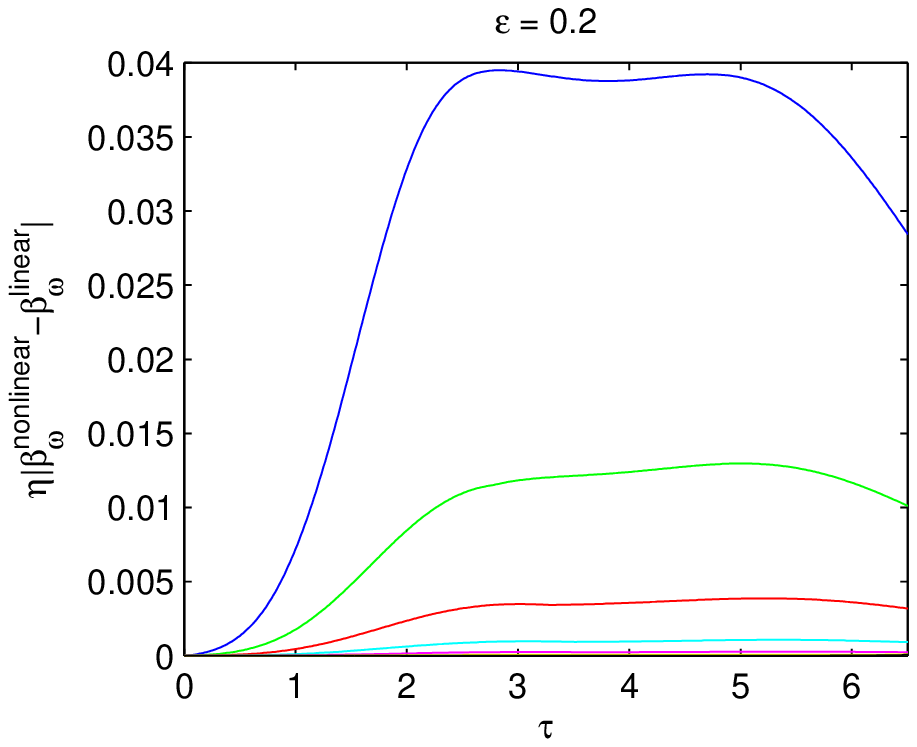}}
\subfigure[]{\includegraphics[scale=0.6]{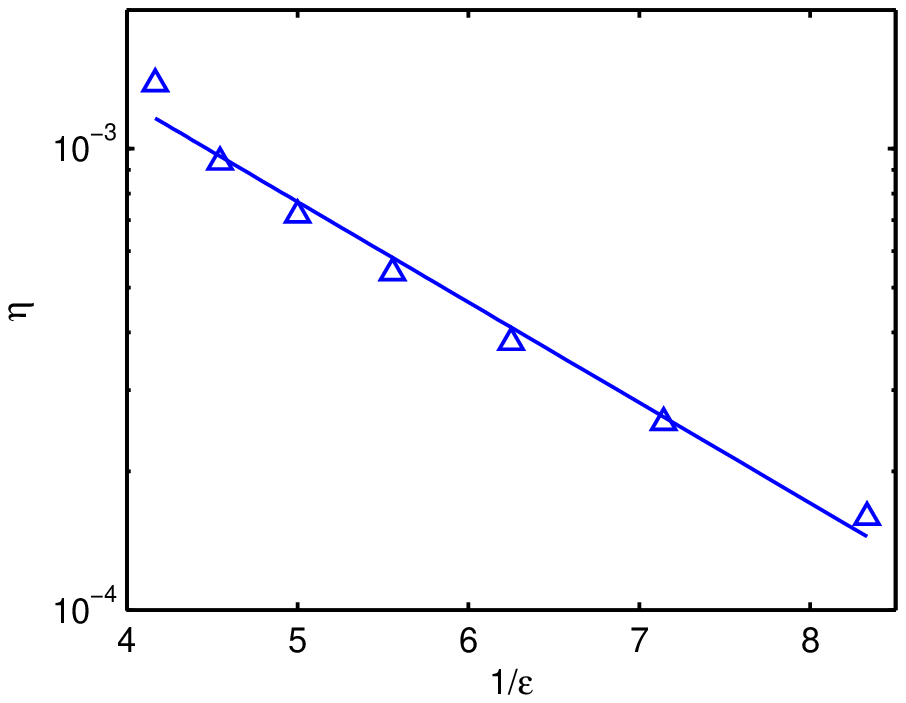}}
\caption{(a) $|\partial_\omega \beta_{nl}-\partial_\omega\beta|$ as a function of $\tau$ for $\eta=0.001 \cdot 2^{-n}$, $n=0,1,\ldots,7$ ($N=256$, $\Delta t=0.001$). (b) $\ln (\eta)$ as a linear function of $1/\epsilon$. The fitting curve is $\ln (\eta) = -0.5/\epsilon -4.66$. }\label{NewFig}
\end{figure}

To close this subsection we briefly consider the range of validity of the linear approximation. As is evident from Fig.~\ref{fig10N}, for a given initial condition this range strongly depends on $\epsilon$. The results of Subsect.~\ref{strongly} suggest that it might decrease exponentially: $\eta < {\rm const}\; e^{-{\rm const}/\epsilon}$, where the constants might depend on the initial condition. To test this hypothesis, we for the initial condition $\beta(\omega,0)=\eta \omega^{10}$ compared the linearized and the nonlinear evolution for values $\eta=0.001 \cdot 2^{-n},n=0,1,\ldots,5$ and $\epsilon$ in the range $0.12\le\epsilon\le0.24$. We specifically calculated the absolute value of the difference $\partial_\omega
\beta_{nl}-\partial_\omega\beta$ in forward direction $\omega=1$. We choose the derivative since it prominently shows up in the nonlinear equations (\ref{Girasol}), (\ref{Azurit}). The results for $\epsilon=0.2$ are shown in Fig.~\ref{NewFig}(a). We observe that after some initial rise this difference saturates at some $\eta$-dependent plateau, where the plateau value strongly increases with $\eta$. Eventually it decreases again, in agreement with the expectation that for the small perturbations $\eta$, the circle is the asymptotic attractor. Interpolating among the plateau values we now for each $\epsilon$ determined a value $\eta^\ast(\epsilon)$ where the plateau value equals 0.02 at $\tau=4$. Fig.~\ref{NewFig}(b) shows $\ln \eta^\ast$ as function of $1/\epsilon$. As expected, it shows an essentially linear decrease. This supports the hypothesis that the range of validity of the linear approximation, and presumably also the basin of attraction of the circle, decrease exponentially with increasing $1/\epsilon$.

\begin{figure}[h]
\begin{center}
\subfigure[$\epsilon=0$,~$n=0:7$]{\includegraphics[scale=0.6]{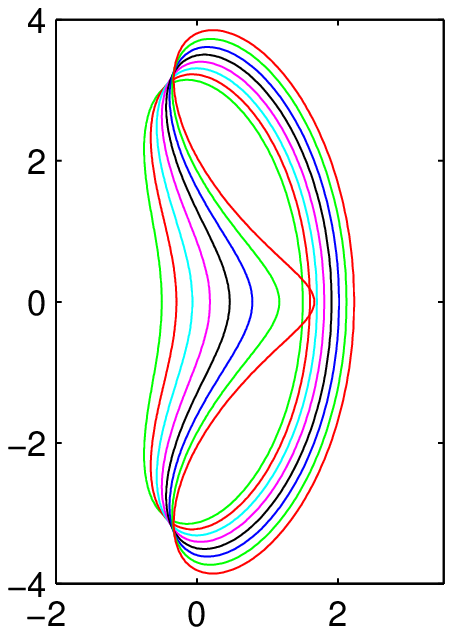}}
\subfigure[$\epsilon=0.01$,~$n{=}0:5$]{\includegraphics[scale=0.6]{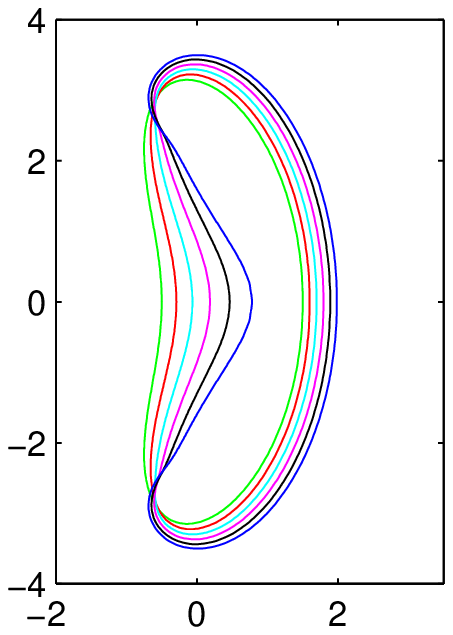}}
\subfigure[$\epsilon=0.1$,~$n=0:10$] {\includegraphics[scale=0.6]{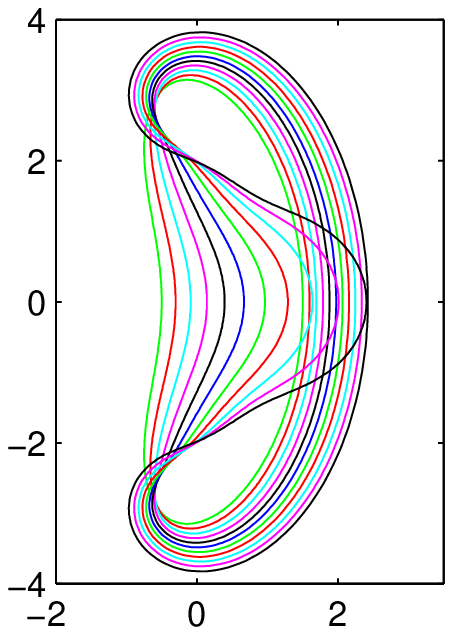}}
\subfigure[$\epsilon=1$,~$n=0:16$]{\includegraphics[scale=0.6]{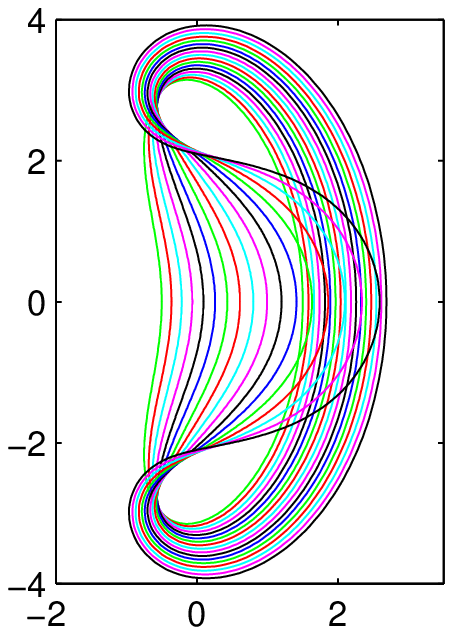}}
\end{center}
\caption{\label{fig:W3} $f(\omega,0)=\frac{1}{\omega}-2\omega+0.5\omega^{2}$ ($N=512$, $\Delta t=0.0005$). The solutions are shown at $t= 0.2n$.} \label{fig13N}
\end{figure}

\begin{figure}[h]
\begin{center}
\subfigure[$\epsilon=0$,~$n=0:5$]{\includegraphics[scale=0.6]{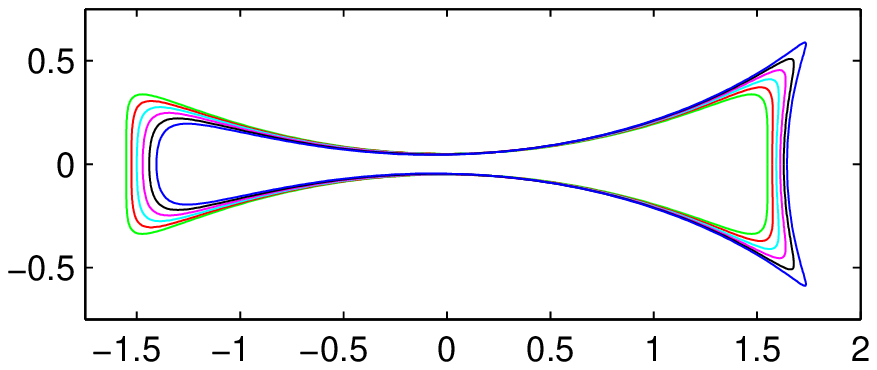}}
\subfigure[$\epsilon=0.01$,~$n=0:4$]{\includegraphics[scale=0.6]{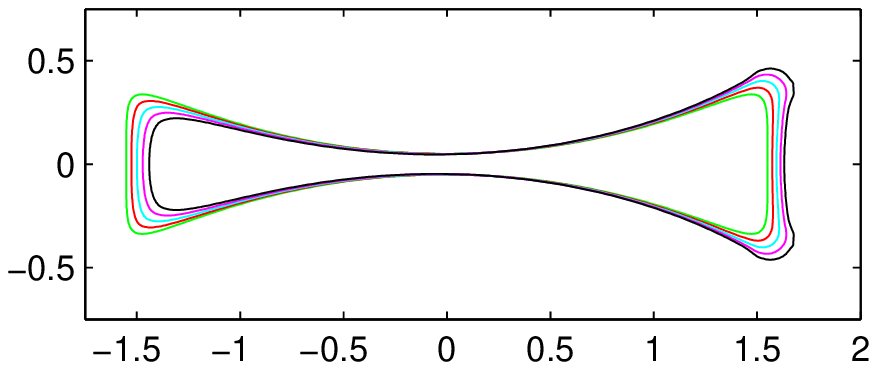}}
\\
\subfigure[$\epsilon=0.1$,~$n=0:10$]{\includegraphics[scale=0.6]{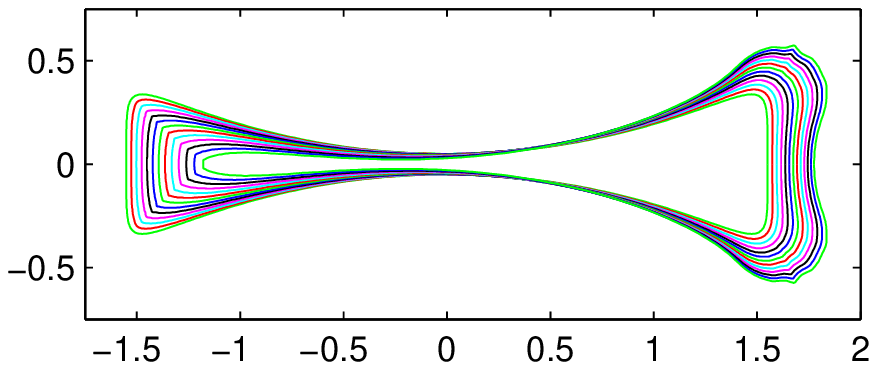}}
\subfigure[$\epsilon=1$,~$n=0:5$]{\includegraphics[scale=0.6]{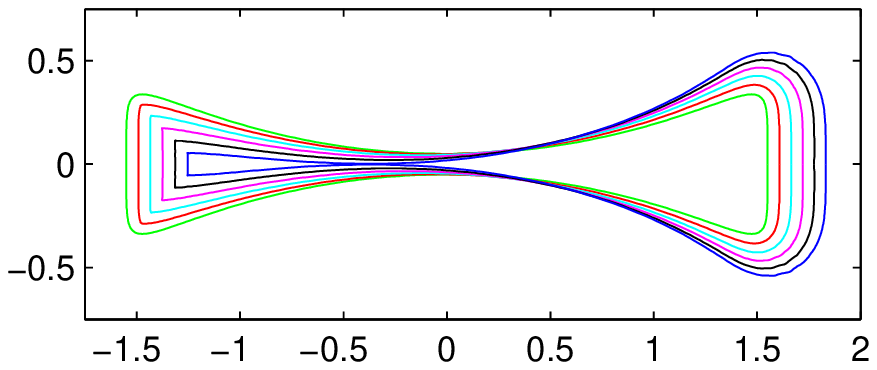}}
\end{center}
\caption{\label{fig:W4} $f(\omega,0)=\frac{1}{\omega}+0.75\omega-0.2\omega^{3}$ ($N=512$, $\Delta t=0.0005$). The solutions are shown at $t= 0.01n$ for (a),(b),(c) and at $t=0.05$ for (d).}
\label{fig14N}
\end{figure}

\begin{figure}[h]
\begin{center}
\subfigure[$\epsilon=0$,~$n=0:25$]{\includegraphics[scale=0.6]{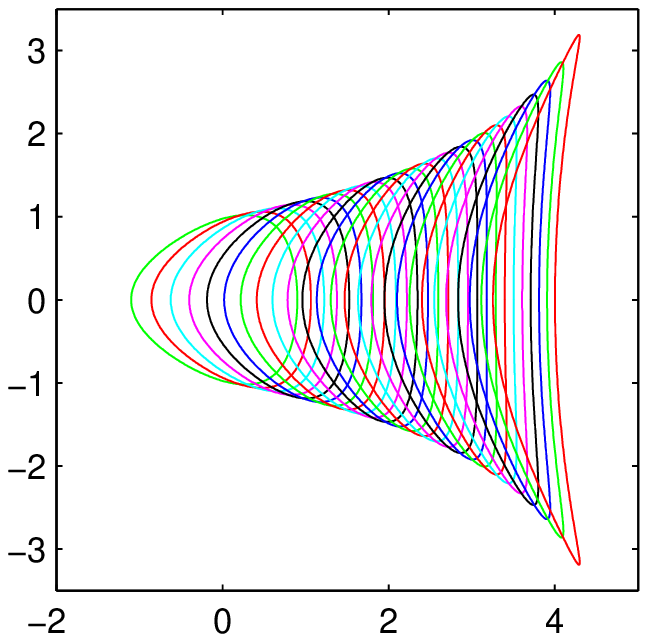}}
\subfigure[$\epsilon=0.01$,~$n=0:6$]{\includegraphics[scale=0.6]{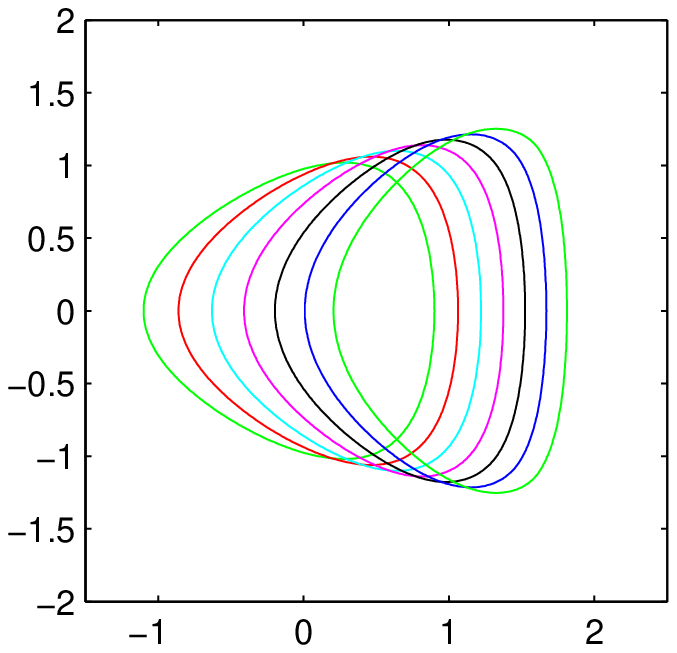}}
\\
\subfigure[$\epsilon=0.1$,~$n=0:12$]{\includegraphics[scale=0.6]{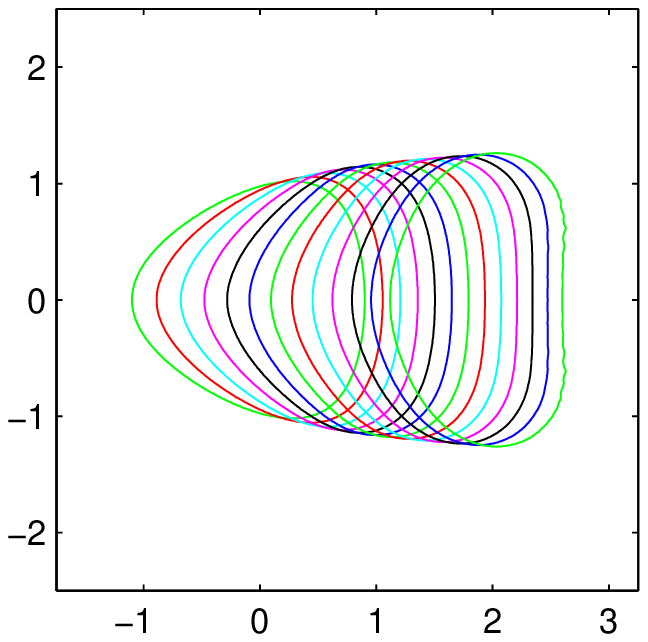}}
\subfigure[$\epsilon=1$,~$n=0:9$]{\includegraphics[scale=0.6]{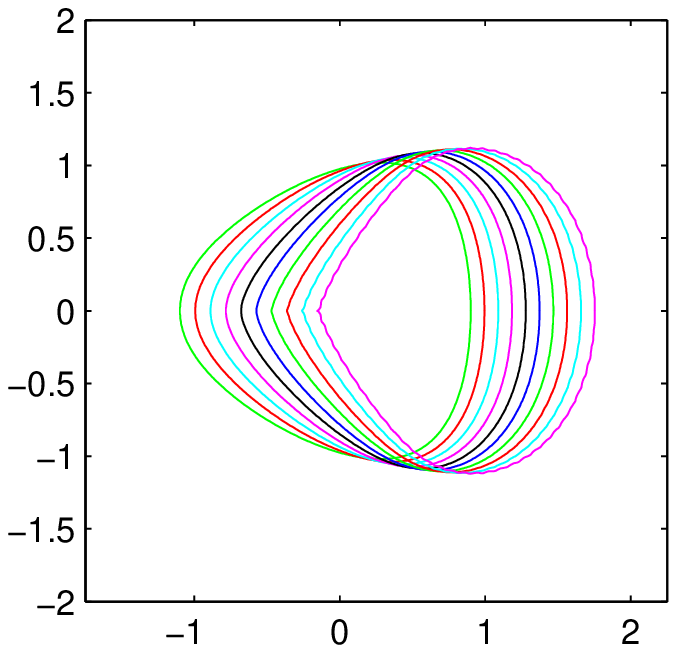}}
\end{center}
\caption{\label{fig:ice_cone}Ice-cone: $f(\omega,0)=\frac{1}{\omega}-0.1\omega^{2}$ ($N=512$, $\Delta t=0.0005$). The solutions are shown at $t= 0.1n$.} \label{fig15N}
\end{figure}

\subsection{Examples of the evolution of general shapes}\label{GenShapes}

For general initial conditions the time evolution may lead to a breakdown of the model by two different
mechanisms. First, a global breakdown occurs at time $t_c$ where the mapping looses the property of being
one-to-one: $f(e^{i \alpha_1},t_c)=f(e^{i \alpha_2},t_c)$, $\alpha_1\neq \alpha_2$. Clearly for $t>t_c$ the
model becomes invalid. Physically we might suspect that the bubble splits into two disjoint parts. Second,
the model can break down locally if a zero of $\partial_{\omega}f(\omega,t)$ reaches the unit circle, which
results in a cusp of the interface. It is well known that this is a common mechanism for breakdown in the
unregularized model, ($\epsilon=0$).

Global breakdown has been observed with curvature regularization (see, e.g., \cite{NieT}), and is also observed in our model. Figs.~\ref{fig13N} and \ref{fig14N} show examples, where each figure shows the evolution of a given initial condition for several values of $\epsilon$. For $\epsilon=0$ cusps do form, as is particularly obvious in Fig.~\ref{fig14N}a. The regularization $\epsilon>0$ suppresses the cusps, but does not change the tendency to split into two parts.

Whether local breakdown by cusp formation can occur in the regularized model, is a more difficult question.
We recall that the neighborhood of $\omega=-1$ shows a special dynamics. The linearized evolution of the
interface can lead to a very complicated shape near $\omega=-1$ since with increasing time singularities of
$f(\omega,0)$ are gathered in this neighborhood. According to Subsect.~\ref{asymptotic} these singularities
for $\epsilon \ge 1/2$ dominate the local structure of the interface. We also note that for
$\epsilon=\infty$ even the linearized evolution is singular at $\omega=-1$, where it produces a spike. It
thus is conceivable that the nonlinear evolution yields a cusp or some other type of singularity at
$\omega=-1$.

We studied this problem with the initial condition $f(\omega,0)=1/\omega-0.1 \,\omega^2$. The results of the
nonlinear evolution are shown in Fig.~\ref{fig15N}. Clearly the cusps forming for $\epsilon=0$ in the front
part are suppressed for $\epsilon>0$. We further observe that for $\epsilon=$1/100 or 1/10, the curvature
near $\omega=-1$ decreases, whereas it increases for $\epsilon=1$. This suggests that for $\epsilon=1$ a cusp may be formed.


\section{Summary and conclusion}

Consistent with the eigenvalue analysis presented in \cite{PartI}, the results of the present paper strongly suggest that a uniformly translating circle is a linearly stable solution of a Laplacian interface model regularized by a kinetic undercooling boundary condition. Furthermore, numerical results of the full nonlinear evolution indicate that the circle has a finite basin of attraction in a space of analytic functions. An important feature of the stabilizing mechanism is the advection of perturbations towards the back of the circle. Except for a small region at the back that asymptotically contracts to a point, the final relaxation to the circle is exponential. With decreasing regularization parameter $\epsilon>0$ the anomalous behavior at the back is suppressed. However, perturbations increase as long as they are in the front half of the circle, and this effect is strongly enhanced by lowering $\epsilon$. Since larger perturbations may lead to branching, this indicates that the basin of attraction of the circle shrinks exponentially with decreasing $\epsilon$.

The interface model considered here is a reduced form of a PDE-model describing the streamer stage of electric breakdown in the simplest physically relevant situation. It ignores the physics inside the streamer and the internal structure of the screening layer; the layer is approximated by the interface together with the boundary condition (\ref{Amethyst}) which introduces the regularization. Numerical solutions of the PDE-model indicate that these approximations for sufficiently strong externally applied fields are justified in the dynamically active front part of the streamer. The back of the streamer is not represented adequately by the interface model. However, the evolution of the streamer and in particular stability or instability against branching is determined by the active head region, which corresponds to the front half of the circle in our analysis. Indeed, numerical solutions of the PDE-model in two dimensions show a behavior quite similar to the evolution of the front half of weakly disturbed circles in the interface model. After
reaching the streamer stage the streamer head is of nearly circular shape and moves with constant velocity. It slowly flattens at the tip and branches. Compared to the results of the interface model as illustrated in Fig.~\ref{fig11N}, the main difference is a slow increase of the head radius due to weak currents flowing into the head from the interior of the streamer.

In summary, we believe that our results not only are of some interest in the context of interface models but also shed some light on the problem of streamer branching.
\\

{\bf Acknowledgements:} S.~Tanveer was supported by US National Science Foundation DMS-0807266 and acknowledges hospitality at CWI Amsterdam. F.~Brau acknowledges a grant of The Netherlands' Organization for Scientific Research NWO within the FOM/EW-program "Dynamics of Patterns". C.-Y.~Kao was partially supported by the National Science Foundation grant DMS-0811003 and an Alfred P.~Sloan Fellowship.

\newpage


\section*{Appendix: Numerical calculation of the nonlinear evolution}

As explained in Sect.~\ref{equations} the shape of the interface is given by
\begin{displaymath} z=x+i y=f\left(e^{i \alpha},t\right), ~~~-\pi<\alpha
\le \pi. \end{displaymath}
We restrict ourselves to interfaces symmetric with respect to the real
axis, so that
\begin{displaymath} f^\ast\left(e^{i \alpha},t\right)=f\left(e^{-i
\alpha},t\right),\end{displaymath}
with the corresponding equation holding for the potential $\Phi
\left(e^{i \alpha},t\right)$. We use the Fourier representation
\begin{eqnarray}f&=&\sum_{k=-1}^\infty a_k(t) e^{i k \alpha},\label{A1}
\\
\Phi&=&\sum_{k=-1}^\infty c_k(t) e^{i k \alpha},\label{A2}
\end{eqnarray}
with a cutoff at high wave number $k=N$. Due to the symmetry, $a_k(t)$
and $c_k(t)$ are real, and the boundary condition at infinity (\ref{Antimonit})
enforces $c_{-1}(t)\equiv a_{-1}(t)$.

For a given shape of the interface the potential is determined by
Eq.~(\ref{Azurit}):
\begin{equation}
|\partial_\alpha f\left(e^{i \alpha},t\right)| ~{\rm Re}\left[ \Phi
\left(e^{i \alpha},t\right)\right]=\epsilon {\rm Re}\left[i \partial_\alpha \Phi
\left(e^{i \alpha},t\right)\right]
\label{A3}
\end{equation}
We represent $|\partial_\alpha f|$ as
\begin{equation}
|\partial_\alpha f|=\sum_{k=-\infty}^\infty d_k(t) e^{i k \alpha},
\label{A4}
\end{equation}
where the symmetry enforces $d_k=d_{-k} \in \mathbb{R}$. For a given $f$ in Fourier representation (\ref{A1}),
\begin{equation}
\partial_\alpha f=\sum_{k=-1}^\infty i k a_k(t) e^{i k \alpha}\label{A1_2}.
\end{equation}
The nonlinear term $|\partial_\alpha f|$ is computed via the standard pseudo-spectral approach, i.e.
$|\partial_\alpha f|$ is obtained in the physical domain via inverse Fourier transform of Fourier coefficients in (\ref{A1_2}) and taking the absolute value and then $d_k$ is determined by the Fourier transform of $|\partial_\alpha f|$. Substituting Eqs.~(\ref{A2}), (\ref{A4}) into Eq.~(\ref{A3}),
we find a system of linear equations for $c_k$, $k \ge 0$, which can be
written as
\begin{equation}\sum_{k=0}^\infty (d_{m-k}+d_{m+k}+\epsilon m\;
\delta_{m,k}) c_k=(\epsilon \delta_{m,1}-d_{m+1}-d_{m-1}) a_{-1},~~~ m \ge 0.
\label{A5}\end{equation}
Here $\delta_{m,k}$ denotes Kronnecker's symbol, and we used the
identity $c_{-1}\equiv a_{-1}$. We solve these equations with a cut off
$k,~m \le N$. Note that $d_{k}$ is needed up to $k=2 N$.

The evolution of the interface is determined by Eq.~(\ref{Girasol}), which can be
written as
\begin{equation}
{\rm Re}\left[\frac{\partial_t f}{\omega\partial_\omega
f}\right]=\frac{{\rm Re}\left[-i \partial_\alpha
\Phi\left(e^{i \alpha}\right)\right]}{|\partial_\alpha f|^2}=R(\alpha).
\label{A6}
\end{equation}
$\frac{\partial_t f}{\omega\partial_\omega f}$ is analytic for $\omega
\in\cal{U}_\omega$ and is real for $\omega=0$ by construction. Eq.~(\ref{A6})
therefore implies
\begin{displaymath}
\left.\frac{\partial_t f(\omega,t)}{\omega\partial_\omega
f(\omega,t)}\right|_{\omega=e^{i \alpha}}=\frac{1}{2 \pi}
\int_{-\pi}^{\pi}\frac{e^{i \alpha'}+\omega}{e^{i \alpha'}-\omega}
R(\alpha')\, d\alpha',
\end{displaymath}
which for $\omega\rightarrow e^{i \alpha}$ reduces to
\begin{equation}\left.\frac{\partial_t f}{\omega\partial_\omega f}\right|
_{\omega=e^{i \alpha}}=R(\alpha)-\frac{i}{2 \pi} P
\int_{-\pi}^{\pi}\cot\frac{\alpha'}{2} ~R(\alpha+\alpha')~ d\alpha', \label{A7}
\end{equation}
where $P$ denotes the principle value. Symmetry enforces
$R(\alpha)=R(-\alpha)$, so that $R(\alpha)$ can be represented as
\begin{equation}
R(\alpha)=\sum_{k=0}^\infty r_k \cos(k \alpha), ~~~r_k\in \mathbb{R},
\label{A8}
\end{equation}
where the $r_k$ again are determined by the
Fourier-cosine transform numerically. Substituting the expansions
(\ref{A1}), (\ref{A8}) into Eq.~(\ref{A7}), we get
\begin{equation}
\frac{d a_k}{dt} = \sum_{n=0}^{k+1} (k-n) a_{k-n} r_n , ~~~k \ge -1.
\label{A9}
\end{equation}
We again truncate this system of ODE's at $k=N$ and solve it via 4-th order Runge-Kutta method (RK4) \cite{RK4}.
Let the initial value problem (\ref{A9}) be specified as follows.
\begin{equation}
\frac{d y}{dt}=f(t,y),\quad y(t_0)=y_0,
\end{equation}
where y denotes the vector function $(a_{-1},a_{0},a_{1},...,a_{N})$.
Then, the RK4 method for this problem is given by the following equations
\begin{eqnarray}
y_{n+1} &=& y_{n}+\frac{1}{6}h (k_1+2 k_2+3 k_3+k_4), \\
t_{n+1} &=& t_{n}+h,
\end{eqnarray}
where $y_{n+1}$ is the RK4 approximation of $y(t_{n+1})$,
\begin{eqnarray}
k_1&=&f(t_n,y_n),\\
k_2&=&f(t_n+\frac{1}{2} h,y_n+\frac{1}{2} h k_1),\\
k_3&=&f(t_n+\frac{1}{2} h,y_n+\frac{1}{2} h k_2),\\
k_4&=&f(t_n+h,y_n+h k_3),
\end{eqnarray}
and $h$ is the time step. In the numerical implementation, $h$ needs to be chosen small enough to ensure numerical stability and it is usually inverse proportional to the cut-ff $N$. The cut-off $N$ needs to be chosen large enough so that the interface can smoothly represented, i.e. the Fourier coefficients are exponentially decayed for large $k$. For most of the initial conditions we used, there are only few Fourier coefficients are not zero. As time evolved, number of nonzero Fourier coefficients will increase. When the high frequency mode is no longer exponentially small, the algorithm needs to be terminated or more Fourier modes need to be used. Adaptive Fourier mode is beyond the scope of this paper. Here we only used fix cut-off $N$ and make sure that the high frequency modes are exponentially small at later time. In the numerical simulations, we use both double and quadruple precision to compute solutions for large enough $N$. To prevent
the spurious growth of the high-wavenumber coefficient generated by run-off error, we filter out the
coefficient which is below the chosen threshold. If the threshold is chosen to be too large, aliasing may occurs. If the threshold is chosen to be too small, it cannot effectively reduce the run-off error. The reasonable choice from experience is about $1000$ bigger than the
run-off error. We choose the threshold to be $10^{-13}$ for double precision and $10^{-29}$ for the
quadruple precision. We can compare the results from both precision to ensure the results we obtained are
not spurious.

Notice that the numerical simulation need to stop at some finite time because of singularity. When singularity is developed, the numerical results become unreliable.
There is a way to test the accuracy of numerical results without knowing the exact solution.
Suppose the numerical method is of $p$-th order, we expect that
\begin{equation}
y^{h}_n-y(nh)= \mathcal{O}(h)^p.
\end{equation}
This implies that
\begin{eqnarray}
y^{h}_n-y^{h/2}_{2n}&=& \mathcal{O}(1-(1/2)^p)(h)^p,\\
y^{h/2}_{2n}-y^{h/4}_{4n}&=& \mathcal{O}(1-(1/2)^p)(h/2)^p.
\end{eqnarray}
Thus the order $p$ can be estimated by using the formula\[
p\approx\log_{2}\left|\frac{y^{h}_n-y^{h/2}_{2n}}{y^{h/2}_{2n}-y^{h/4}_{4n}}\right|.\] We choose the initial condition $f(\omega,0)=\frac{1}{\omega}-0.1\omega^{2}$ and compute solutions for step size $0.001$, $0.0005$, and $0.00025$ with $N=64$. In Fig. \ref{fig:ice_cone}, we can see that the order stays close to $4$ up to the time equals to $2.5$, $0.7$, $1.25$, and $0.9$ for
(a) $\epsilon$=0 (b) $\epsilon$=0.01 (c) $\epsilon$=0.1 and (d) $\epsilon=1$ respectively. Another way to test the accuracy is to check whether the area conservation holds or break down. The area enclosed by the interface should remain as a constant and it can be estimated by
$$
A = - \sum_{k=-1}^{N} k a_k(t)^2.
$$
In Fig. \ref{fig:ice_cone_area}, the area changes versus time are shown for step size $0.00025$. It is clearly that the area conservation and order of accuracy break down at the similar time for (a) $\epsilon$=0, (b) $\epsilon$=0.01, and (c) $\epsilon$=0.1. Notice that, in accuracy test, the order drops to the first order at $t\approx 1$ and the area conservation still holds up to $t\approx 1.3$. Similar behaviors are observed for other initial conditions and different Fourier modes $N$. For the computational results shown in the manuscipt, we show the sulotions up to the time that acuuracy of the solutions can be assured.

\begin{figure}[H]
\begin{center}
\includegraphics[scale=0.5]{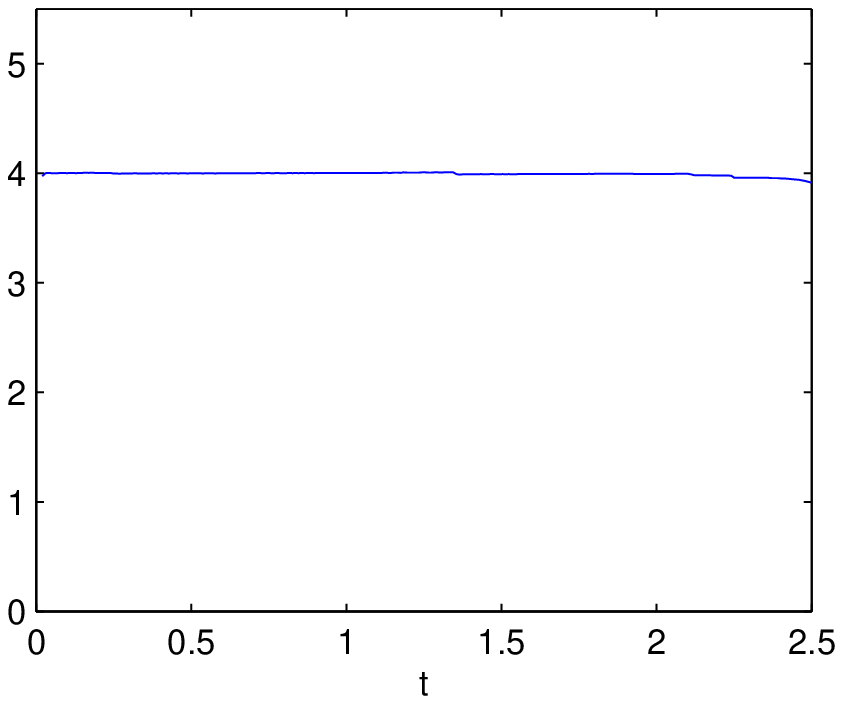}\includegraphics[scale=0.5]{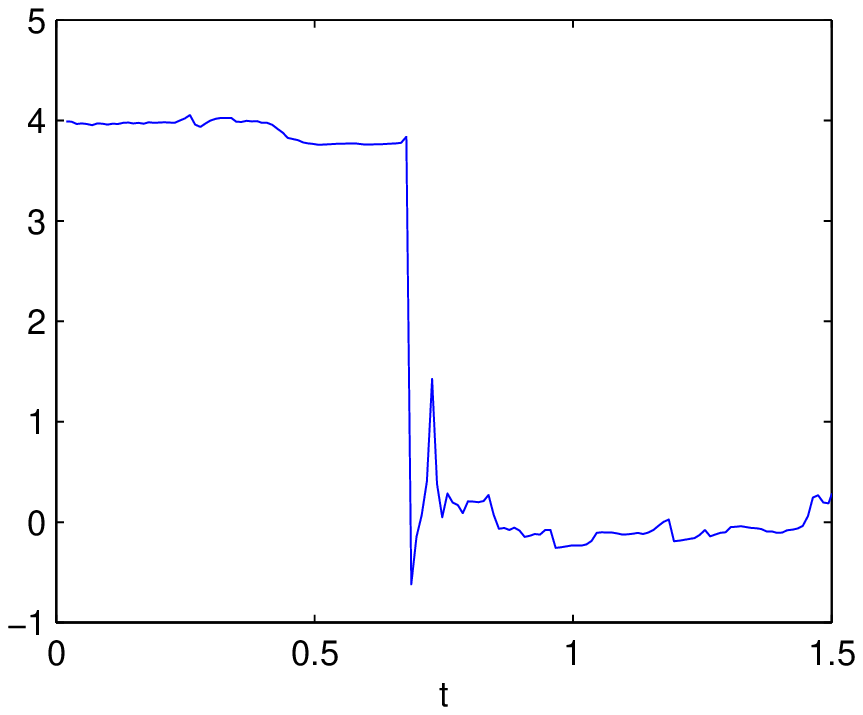}
\par\end{center}

\begin{center}
\includegraphics[scale=0.5]{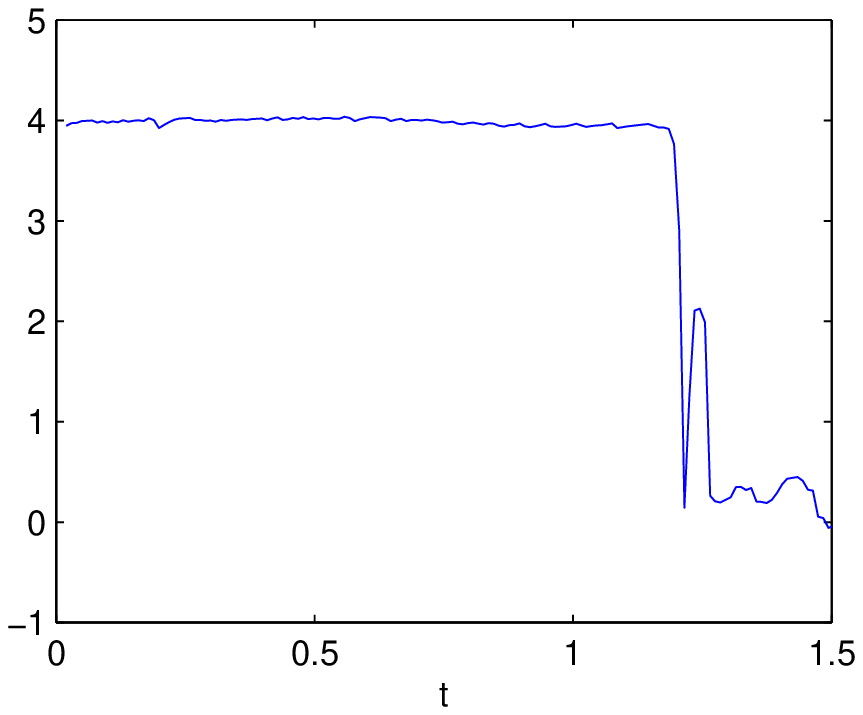}\includegraphics[scale=0.5]{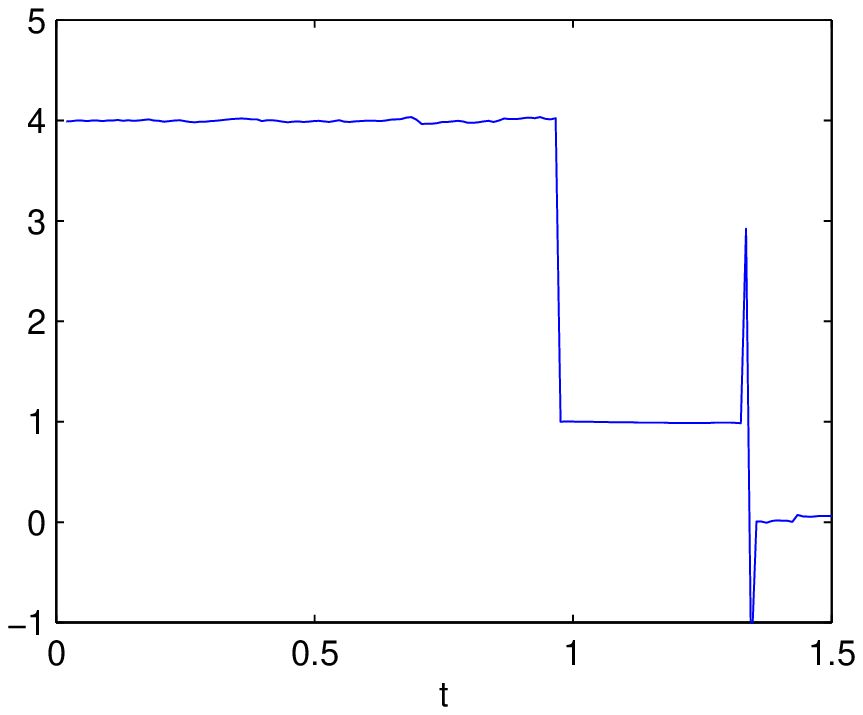}
\par\end{center}

\caption{\label{fig:ice_cone} Numerical estimation of the order of accuracy for the initial condition $f(\omega,0)=\frac{1}{\omega}-0.1\omega^{2}$ with (a) $\epsilon$=0, (b) $\epsilon$=0.01, (c) $\epsilon$=0.1, and (d) $\epsilon$=1. ($N=64$)}

\end{figure}

\begin{figure}[H]
\begin{center}
\includegraphics[scale=0.5]{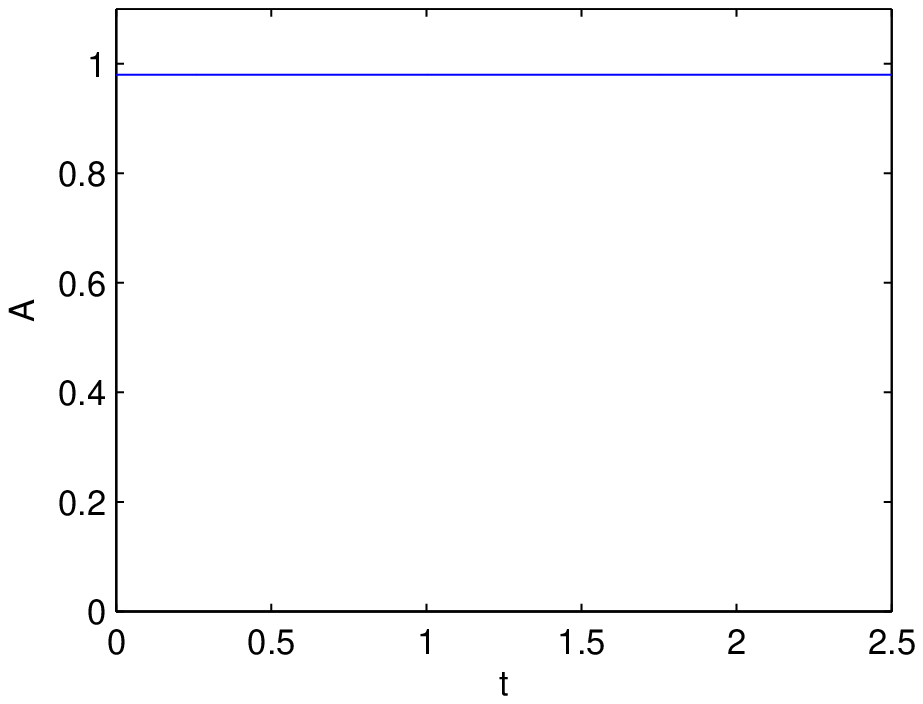}\includegraphics[scale=0.5]{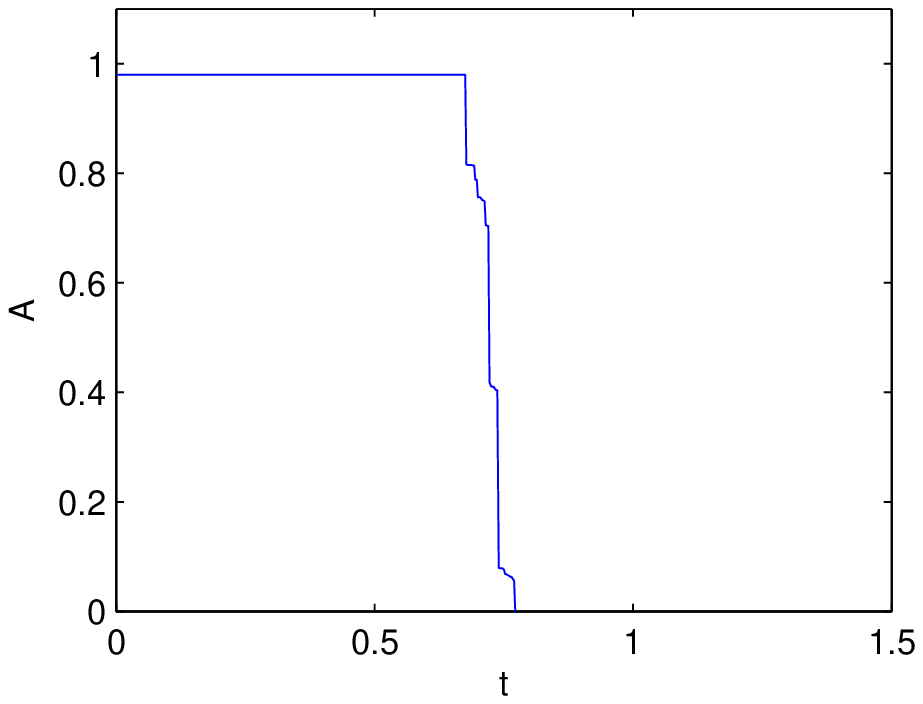}
\par\end{center}

\begin{center}
\includegraphics[scale=0.5]{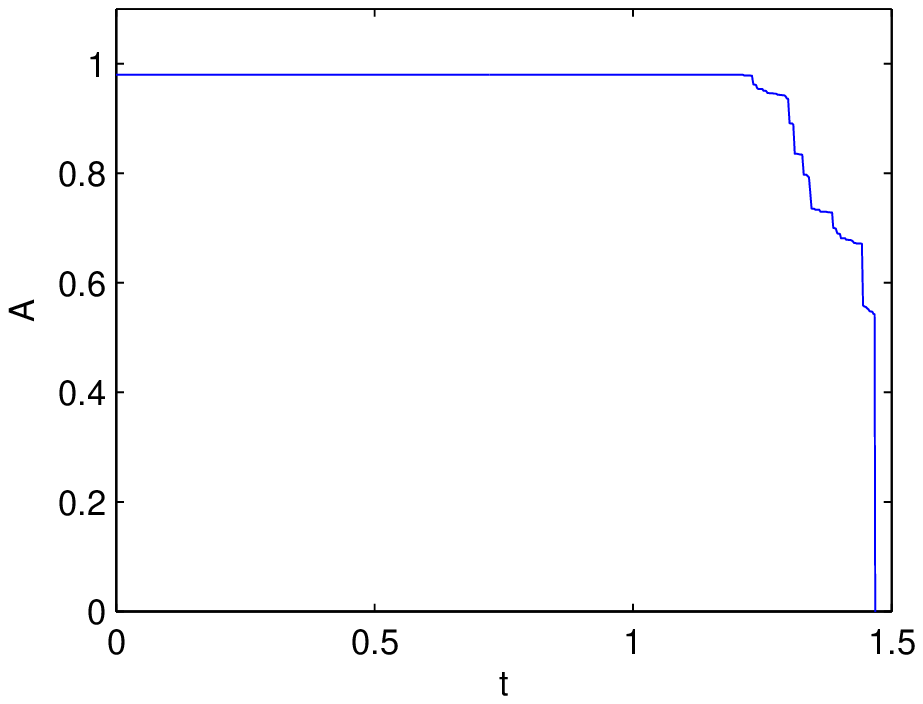}\includegraphics[scale=0.5]{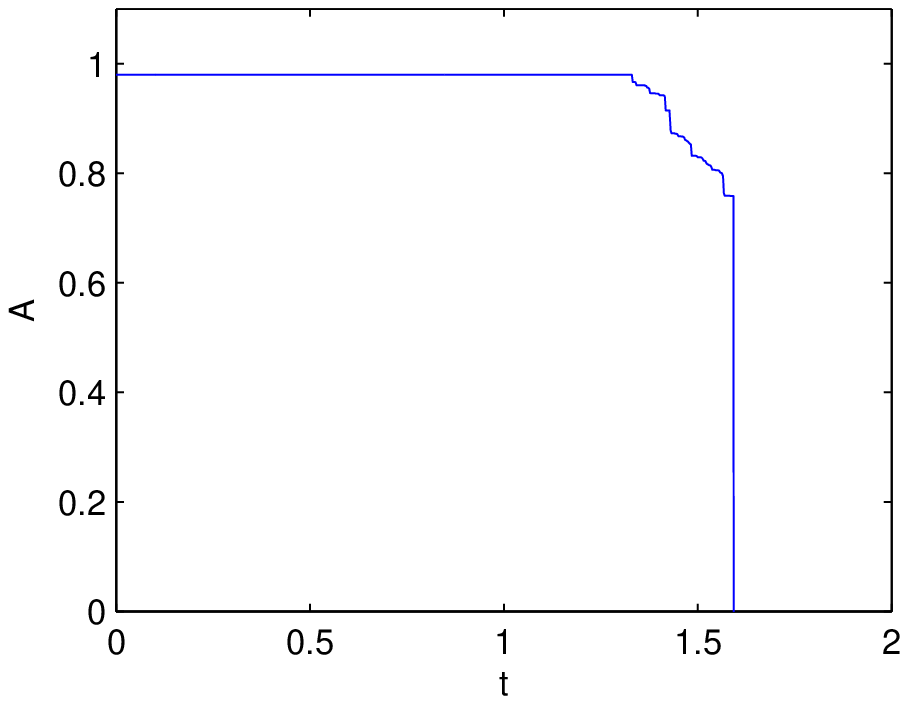}
\par\end{center}

\caption{\label{fig:ice_cone_area} Estimation of area for the initial condition $f(\omega,0)=\frac{1}{\omega}-0.1\omega^{2}$ with (a) $\epsilon$=0, (b) $\epsilon$=0.01, (c) $\epsilon$=0.1, and (d) $\epsilon$=1. ($N=64$)}

\end{figure}

\newpage

\end{document}